\documentclass[aps,pra,reprint,superscriptaddress,floatfix]{revtex4-2}

\usepackage{graphicx}
\usepackage{amsmath}
\usepackage{amssymb}
\usepackage{xcolor}
\usepackage[normalem]{ulem}
\usepackage{hyperref}

\usepackage{booktabs}
\usepackage{multirow}
\usepackage{tabularx}
\usepackage{longtable}
\usepackage{algpseudocode}

\newcounter{algorithm}
\renewcommand{\thealgorithm}{S\arabic{algorithm}}

\hypersetup{
  hidelinks,
  pdftitle={fault-tolerant quantum computing architecture},
  pdfauthor={Yunxin Tang et al.},
  pdfsubject={Fault-tolerant quantum-computing resource estimation}
}

\begin{document}

\title{Low-cost algorithm-to-execution framework for surface-code quantum computing}

\author{Yunxin Tang}
\thanks{These authors contributed equally to this work.}
\affiliation{Center on Frontiers of Computing Studies, Peking University, Beijing 100871, China}
\affiliation{School of Computer Science, Peking University, Beijing 100871, China}
\author{Zixuan Huo}
\thanks{These authors contributed equally to this work.}
\affiliation{Center on Frontiers of Computing Studies, Peking University, Beijing 100871, China}
\affiliation{School of Computer Science, Peking University, Beijing 100871, China}
\author{Junxiang Huang}
\thanks{These authors contributed equally to this work.}
\affiliation{Center on Frontiers of Computing Studies, Peking University, Beijing 100871, China}
\affiliation{School of Computer Science, Peking University, Beijing 100871, China}
\author{Zhou You}
\affiliation{Center on Frontiers of Computing Studies, Peking University, Beijing 100871, China}
\affiliation{School of Computer Science, Peking University, Beijing 100871, China}
\author{Zhirao Wang}
\affiliation{Center on Frontiers of Computing Studies, Peking University, Beijing 100871, China}
\affiliation{School of Computer Science, Peking University, Beijing 100871, China}
\author{Zhenai Ding}
\affiliation{Center on Frontiers of Computing Studies, Peking University, Beijing 100871, China}
\affiliation{School of Computer Science, Peking University, Beijing 100871, China}
\author{Yumeng Zeng}
\affiliation{Center on Frontiers of Computing Studies, Peking University, Beijing 100871, China}
\affiliation{School of Computer Science, Peking University, Beijing 100871, China}
\author{Yangyu Lu}
\affiliation{Center on Frontiers of Computing Studies, Peking University, Beijing 100871, China}
\affiliation{School of Computer Science, Peking University, Beijing 100871, China}
\author{Yiming Huang}
\affiliation{Institute of High Energy Physics, Chinese Academy of Sciences, Beijing 100049, China}
\affiliation{China Center of Advanced Science and Technology, Beijing 100190, China}
\affiliation{Center on Frontiers of Computing Studies, Peking University, Beijing 100871, China}
\affiliation{School of Computer Science, Peking University, Beijing 100871, China}
\author{Zongkang Zhang}
\affiliation{Hefei National Research Center for Physical Sciences at the Microscale and School of Physical Sciences, University of Science and Technology of China, Hefei 230026, China}
\affiliation{Shanghai Research Center for Quantum Science and CAS Center for Excellence in Quantum Information and Quantum Physics, University of Science and Technology of China, Shanghai 201315, China}
\affiliation{Hefei National Laboratory, University of Science and Technology of China, Hefei 230088, China}
\author{Haipeng Xie}
\affiliation{Graduate School of China Academy of Engineering Physics, Beijing 100193, China}
\author{Ruipeng Xing}
\affiliation{Beijing Yuanjie Exploration Quantum Technology Co., Ltd., Beijing 100800, China}
\author{Qiming Ding}
\affiliation{Center on Frontiers of Computing Studies, Peking University, Beijing 100871, China}
\affiliation{School of Computer Science, Peking University, Beijing 100871, China}
\author{Ying Li}
\affiliation{Graduate School of China Academy of Engineering Physics, Beijing 100193, China}
\author{Jinzhao Sun}
\affiliation{School of Physical and Chemical Sciences, Queen Mary University of London, London E1 4NS, United Kingdom}
\author{Xiao Yuan}
\affiliation{Center on Frontiers of Computing Studies, Peking University, Beijing 100871, China}
\affiliation{School of Computer Science, Peking University, Beijing 100871, China}
\author{Yuan Yao}
\affiliation{Center on Frontiers of Computing Studies, Peking University, Beijing 100871, China}
\affiliation{School of Computer Science, Peking University, Beijing 100871, China}

\date{October 14, 2026}


\begin{abstract}
Fault-tolerant quantum computing is progressing from demonstrations of individual logical components toward the execution of useful quantum algorithms on large-scale error-corrected processors. Realizing this transition requires co-design across algorithms, fault-tolerant implementations, and hardware resources. Here we develop an algorithm-to-execution framework for low-cost surface-code quantum computing that carries the structure of a computation from hierarchical logical descriptions to circuit-specific spatial organization, non-Clifford resource planning, and physical resource evaluation. We apply the framework to twenty benchmark circuits spanning seven algorithm families and to a hierarchically reconstructed application-scale elliptic-curve discrete-logarithm workload. Under the stated error and resource-supply models, combining circuit-specific layouts with hybrid non-Clifford implementation reduces space--time volume by up to $12.70\times$ relative to an all-synthesis Litinski fast-block baseline. Circuit-specific layouts reduce both modeled runtime and space--time volume in eighteen benchmarks, while the remaining two retain the fast layout. These results show that the physical cost of fault-tolerant computation emerges from the interaction between algorithm structure, error-corrected implementation, and hardware organization rather than from aggregate logical resource counts alone.

\end{abstract}

\maketitle

\section{Introduction}

Fault-tolerant quantum computing is approaching a transition from demonstrating individual error-corrected components to executing useful quantum algorithms on large-scale quantum processors. Recent experiments have demonstrated substantial progress toward fault-tolerant quantum computation across several hardware platforms. Superconducting processors have demonstrated scalable surface-code and color-code error correction, including below-threshold operation, together with encoded magic-state preparation~\cite{GoogleNature2023surface,GoogleNature2025belowthreshold,Tan2025belowthreshold,Lacroix2025Scaling,Gupta2024Encoding}. Trapped-ion processors have demonstrated repeated quantum error correction, fault-tolerant logical gate sets, logical teleportation, and code switching~\cite{postler2022demonstration,ryan2024high,wang2026,Lin2026logicaloperations}. Neutral-atom systems have progressed from reconfigurable logical processors to repeated surface-code error correction, fault-tolerant logical operations, and logical magic-state distillation~\cite{Bluvstein2023logicalprocessor,BluvsteinNature2025FTQC,Bluvstein2026fault,SalesRodriguez2025Experimental}. Photonic architectures have meanwhile demonstrated substantial progress in scaling and modular networking toward fault-tolerant quantum processors~\cite{aghaee2025scaling,Wang2026Demonstration}. In parallel, theoretical and architectural advances have continued to reduce the overhead of fault-tolerant resource-state preparation and logical execution~\cite{gidney2024magicstatecultivationgrowing,Zhang2025bivariate}. These advances increasingly shift the central systems question from whether fault-tolerant logical components can be realized to how complete quantum computations can be executed with the available error-corrected hardware.

The resources required for application-scale fault-tolerant quantum computation have been studied for many years. For cryptographic applications, detailed estimates have been developed for both integer factorization~\cite{Gidney2021RSA,Gidney2025RSA} and elliptic-curve discrete logarithms~\cite{proos2003shor,roetteler2017quantum,haener2020improved}, translating algorithmic constructions into logical gate counts, qubit requirements, and ultimately physical execution costs. For quantum simulation and quantum chemistry, related studies have quantified the logical and fault-tolerant resources required for Hamiltonian simulation, ground-state energy estimation, and chemically relevant systems~\cite{Babbush2018Encoding,su2021fault,lee2021even,kim2022fault,goings2022reliably}. These studies established quantitative targets for fault-tolerant quantum computing and showed that algorithmic improvements, logical compilation, code parameters, and resource-state preparation can substantially change the physical requirements of a computation~\cite{Litinski2019gameofsurfacecodes,Beverland2022assessing,Gidney2025RSA,Babbush2026Securing,toshio2025practical,Toshio2026starmagicmutation}. Related studies have also explored code- and hardware-specific resource models for qLDPC codes and architectures with nonlocal or reconfigurable connectivity~\cite{Bravyi2024highthresholdBB,Xu2024Constant,ZhangLi2025qldpc,Mathews2026Placing,Yang2026hardwareTailored}.

More recent resource models have moved closer to physical execution by incorporating placement and routing, lattice-surgery scheduling, reaction-time constraints, communication resources, and magic-state generation~\cite{Kan2025sparo,Huggins2025flasq,Hofmeyr2025PureMagic,zhou2026topols,pflieger2026harvest,tornow2026rushhour}. As resource estimates move closer to physical execution, a broader issue becomes apparent: the cost of fault-tolerant computation is jointly determined by the algorithm, its fault-tolerant realization, and the organization of the underlying hardware. Algorithmic interaction and parallelism shape communication and spatial demand, while fault-tolerant choices determine ancillary and non-Clifford resources whose physical placement further affects routing and execution time. Resource estimation therefore becomes a cross-layer design problem rather than a simple accumulation of independent overheads.

Here we investigate this cross-layer problem at the execution level of surface-code quantum computing. Rather than mapping a logical circuit onto a predefined fault-tolerant architecture, we carry the structure of the computation through the implementation process and use it to organize the resources required for execution. Starting from hierarchical algorithm descriptions, we construct dependency-preserving logical schedules and extract an executable-workload representation that retains logical interaction structure, operation parallelism, and time-resolved non-Clifford demand. These quantities are then used to guide circuit-specific surface-code organization and non-Clifford resource planning, followed by physical resource evaluation under a common error and timing model. The resulting algorithm-to-execution framework therefore connects algorithmic structure, fault-tolerant realization, and hardware resource organization within a single traceable workflow.

We evaluate the framework on twenty benchmark circuits spanning seven algorithm families, twelve additional QAOA parameter-checkpoint instances, and a hierarchically reconstructed application-scale elliptic-curve discrete-logarithm workload. Under the stated error and resource-supply models, combining circuit-specific surface-code layouts with hybrid non-Clifford implementation reduces space--time volume by up to $12.70\times$ relative to an all-synthesis Litinski fast-block baseline. With a common lattice-surgery scheduler and resource-access convention, the layout search reduces modeled execution time and space--time volume in eighteen Hybrid benchmarks and retains fast layouts for the remaining two. The application-scale elliptic-curve case further demonstrates how the same algorithm-to-execution methodology can be applied through hierarchical composition when full circuit expansion is impractical. Together, these results show that the cost of fault-tolerant quantum computation is not fixed by aggregate logical resource counts alone, but emerges from how the algorithm, fault-tolerant implementation, and physical architecture are coordinated during execution.

\section{Results}

\subsection{From logical computation to executable surface-code requirements}

\begin{figure*}
    \centering
    \includegraphics[width=1.0\linewidth]{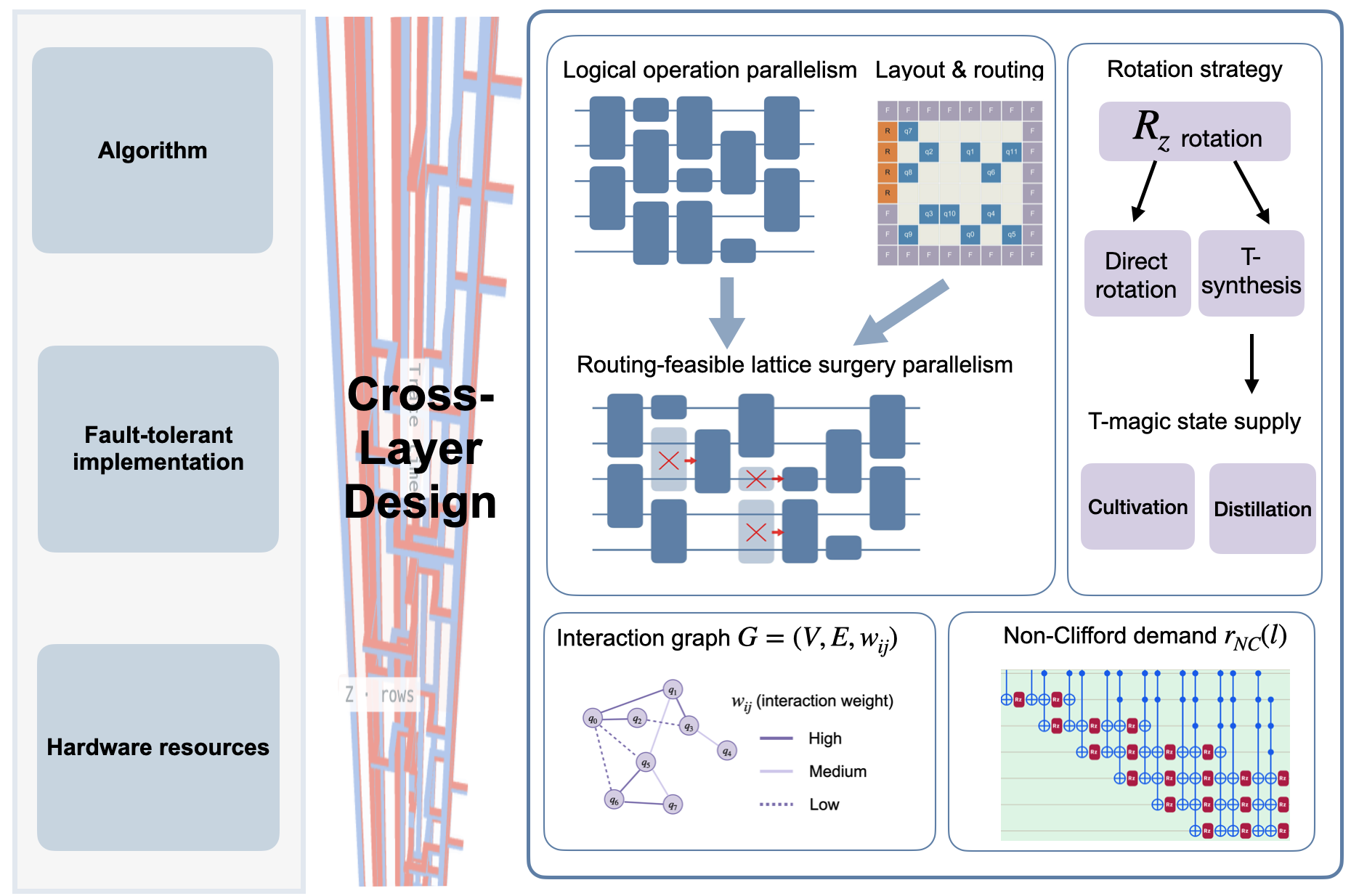}
\caption{\textbf{Cross-layer design from quantum algorithms to fault-tolerant execution.} Algorithm structure is carried through logical compilation, patch-level fault-tolerant realization, and hardware resource organization rather than being reduced to aggregate resource counts. The central space--time representation illustrates the intermediate logical-patch level, where scheduled logical operations are translated into surface-code activity over space and time. Information retained from the computation is then used to coordinate physical organization and execution: logical parallelism is reconciled with routing constraints, interaction structure informs communication and layout, and non-Clifford requirements determine the corresponding rotation and resource-state strategies. Together, these layers form the algorithm-to-execution interface used for cross-layer resource optimization.}
    \label{fig:fig1}
\end{figure*}

A low-cost fault-tolerant implementation cannot be determined from logical resource counts alone. Logical qubit count, circuit depth, and non-Clifford gate count provide useful measures of algorithmic complexity, but they do not specify how a computation should be organized and executed on a fault-tolerant processor. Logical interaction structure, operation parallelism, and the temporal demand for non-Clifford resources directly affect communication, spatial organization, resource-state supply, and ultimately the physical footprint and runtime of the implementation.

Existing resource-estimation and compilation approaches often evaluate a logical circuit on a predefined execution substrate~\cite{Litinski2019gameofsurfacecodes, Gidney2025RSA, Yang2026hardwareTailored}. This abstraction can hide the fact that computations with similar aggregate logical resources may impose different implementation requirements once their execution structure is taken into account. A circuit with dense logical interactions may require substantial communication resources, whereas another with strongly time-dependent non-Clifford demand may require a different resource-state supply strategy. Preserving this information is therefore necessary before the subsequent surface-code implementation can be determined.

To carry this information through the algorithm-to-execution workflow, we introduce the executable workload as an intermediate representation. We denote it by $\mathcal{W}$ and retain the three execution descriptors used in the subsequent optimization,
\begin{equation}
\mathcal{W}
=
\left\{
G_{\mathrm{int}},
P(\ell),
r_{\mathrm{NC}}(\ell)
\right\},
\label{eq:workload_definition}
\end{equation}
where $G_{\mathrm{int}}$ denotes the weighted logical interaction structure, $P(\ell)$ records the number of operations assigned to logical layer $\ell$, and $r_{\mathrm{NC}}(\ell)$ records the non-Clifford demand in that layer. These descriptors summarize the workload; the ordered circuit, qubit supports, boundary axes, rotation angles, and available measurement metadata remain attached to the handoff and are required to reconstruct execution.

We construct the executable workload by carrying the structure of the algorithm into a dependency-preserving logical execution. Hierarchical algorithm descriptions are first decomposed into reusable computational modules and logical operations, which are then scheduled while preserving quantum and classical dependencies, measurements, resets, and rotation-angle information. Rather than reducing the computation to aggregate gate counts, we retain the execution structure required for subsequent fault-tolerant realization. As illustrated in Fig.~\ref{fig:fig1}, this information is propagated through an intermediate logical-patch level, where the temporal structure of the computation is expressed as surface-code space--time activity and constrained by patch accessibility and routing. It is then used to determine the spatial organization and non-Clifford resource strategy of the physical implementation. The executable workload therefore provides the interface through which algorithm structure is carried across logical compilation, fault-tolerant execution, and hardware resource organization.

The extracted workloads expose execution requirements that are hidden by conventional logical metrics. Across the representative circuits considered here, the weighted interaction graphs, scheduled parallelism, and temporal non-Clifford demand vary substantially even among circuits with similar logical-scale descriptions. Arithmetic workloads tend to generate complex interaction patterns and long dependency chains, whereas variational and simulation workloads can exhibit greater parallelism and strongly non-uniform non-Clifford demand. These differences motivate the circuit-specific spatial and resource-supply decisions considered in the following sections.

The same representation extends to application-scale algorithms whose complete elementary circuits are too large to materialize and schedule explicitly. For the secp256k1 elliptic-curve discrete-logarithm computation of Luo \textit{et al.}~\cite{luo2026quantumalgorithmellipticcurve}, we distinguish source-reconstructed circuit modules from the paper-level algorithmic composition. The strongest source-reconstructed resource-counting boundary is one controlled affine point addition. Below this boundary, the prime-field arithmetic, extended-Euclidean-algorithm (EEA) subroutines, elementary gates, measurements, resets, and classical feed-forward dependencies are reconstructed from the released implementation. Above it, the signed-window structure, point-addition multiplicity, table-lookup cost, and semiclassical-QFT structure follow the paper-level construction. Resources are recursively composed according to module invocation multiplicity without materializing the complete elementary circuit. The resulting hierarchy combines source-reconstructed point-addition modules with paper-level algorithmic multiplicities and is evaluated by composing selected local LS profiles with area-constrained family schedules and a shared finite T-state supply.

The executable workload therefore provides the intermediate representation that connects logical computation to the subsequent stages of the algorithm-to-execution framework. In the following sections, its execution requirements are used to synthesize circuit-specific surface-code layouts, select non-Clifford resource-supply strategies, and evaluate the resulting fault-tolerant implementation.

\subsection{Synthesizing low-cost surface-code architectures from execution requirements}

Large-scale surface-code architectures are often designed as general-purpose execution substrates, with logical data blocks, communication regions, and non-Clifford resource interfaces arranged to support a broad range of computations. Such layouts provide flexibility, but they do not exploit the spatial and temporal structure of a specific scheduled computation. The executable workload introduced above makes this structure explicit: logical interaction patterns determine communication demand, operation parallelism constrains simultaneous access to routing resources, and time-resolved non-Clifford demand affects the required accessibility to resource-state interfaces.

We therefore treat the surface-code layout as part of the algorithm-to-execution process rather than as a fixed backend. For a given executable workload $\mathcal{W}$, the architecture module searches a cost function over candidate layouts,
\begin{equation}
\mathcal{A}^{*}
=
\arg\min_{\mathcal{A}}
C(\mathcal{A}|\mathcal{W}),
\label{eq:workload_architecture}
\end{equation}
where $\mathcal{A}$ denotes a candidate surface-code architecture and $C(\mathcal{A}|\mathcal{W})$ accounts for interaction-weighted communication distance, resource-state access, and logical-boundary compatibility. The objective incorporates communication and boundary access rather than geometric distance alone. In the resource experiments, the returned layouts are subsequently evaluated with their executable schedules and resource fleets; the lowest architecture proxy and lowest physical space--time volume need not select the same candidate.

\begin{figure*}[htbp]
    \centering
    \begin{minipage}[t]{0.95\textwidth}
        \includegraphics[width=0.95\linewidth]{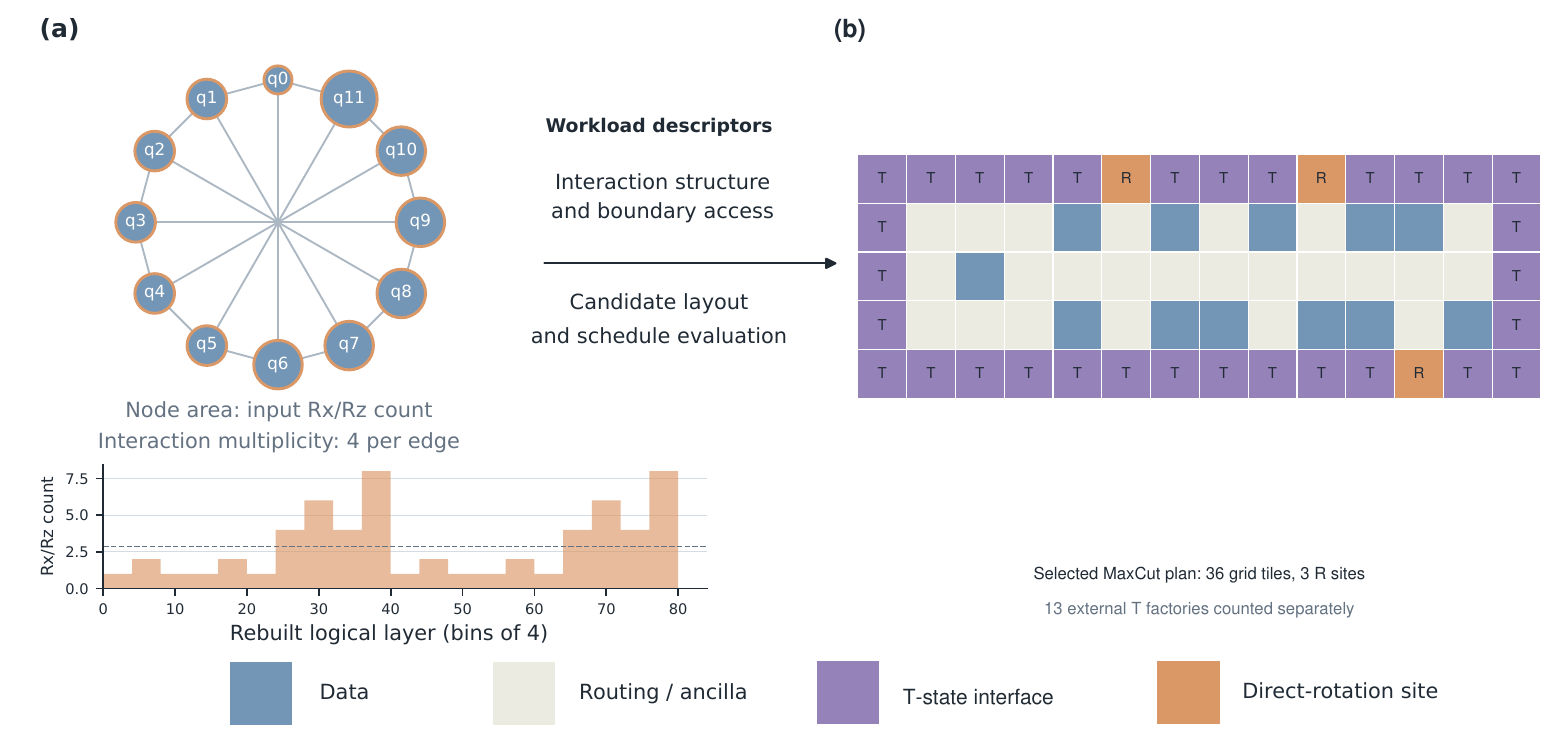}
    \end{minipage}
    \par\medskip
    \begin{minipage}[c]{0.9\textwidth}
        \includegraphics[width=\linewidth]{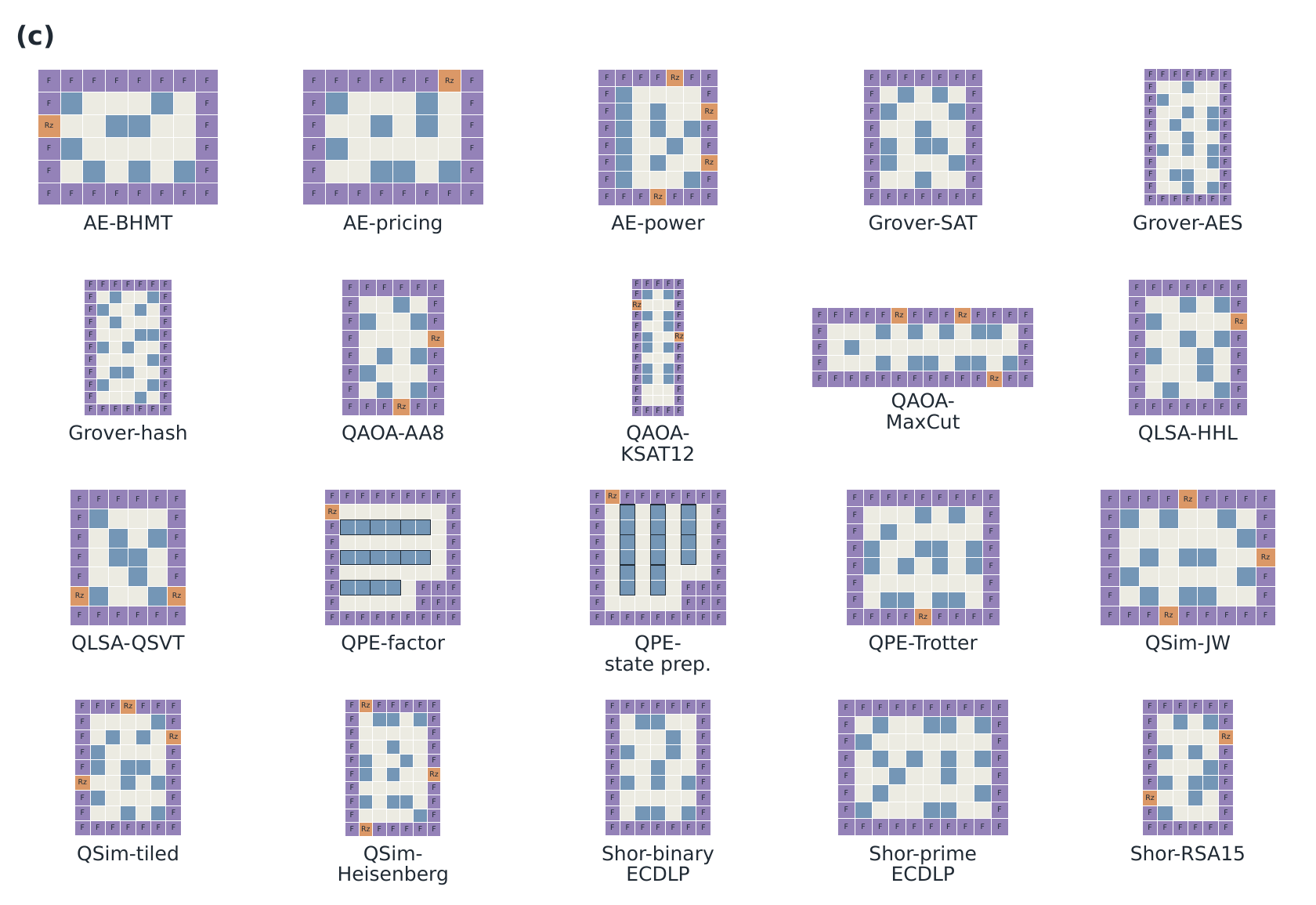}
    \end{minipage}

\caption{\textbf{Circuit-specific surface-code architecture synthesis from execution requirements.} \textbf{(a)} Interactions and input $R_x/R_z$ demand of the twelve-qubit QAOA--MaxCut instance. Node area is proportional to rotation count; all eighteen interaction edges have multiplicity four. The lower trace sums rotations in consecutive groups of four rebuilt logical layers. \textbf{(b)} The saved selected Hybrid layout: 36 data/routing tiles, three direct-rotation sites, and thirteen external T factories charged separately. Perimeter T labels denote state-access interfaces, not factory bodies. Per-qubit rotation demand is a descriptor, not an independently verified placement objective. \textbf{(c)} Selected Hybrid layouts for all twenty benchmarks, in benchmark-table order. QPE--factor and QPE--state preparation retain fast layouts; QPE--Trotter selects a circuit-specific layout. Colors follow panel (b); F denotes the T-state supply boundary and Rz denotes direct-rotation access. Fast layouts have a closed schematic supply perimeter, including corners and recessed edge positions; completing this display does not add modeled ports, factory bodies, or charged area. Outlines identify shared two-qubit patches. Layouts are not drawn to a common physical scale.}
    \label{fig:architecture}
\end{figure*}

Figure~\ref{fig:architecture} illustrates how workload information is translated into a physical layout. For the complete 12-qubit QAOA--MaxCut $p=2$ logical representation in Fig.~\ref{fig:architecture}(a), a displayed edge denotes a direct two-qubit interaction, whereas the absence of an edge denotes no direct interaction. The circular drawing is a schematic visualization rather than a hardware topology. The node area records the per-qubit $R_x/R_z$ count, and the lower trace aggregates the scheduled demand over consecutive groups of four logical layers for visual clarity. These descriptors are related but serve different stages: the interaction structure guides the circuit-specific placement and routing search, while the layer-resolved non-Clifford demand is used in subsequent resource-supply evaluation. The resulting grid in Fig.~\ref{fig:architecture}(b) therefore need not place the qubit with the largest $R_x/R_z$ count closest to a displayed direct-rotation access site, because per-qubit $R_x/R_z$ proximity is not a separate placement objective in the present model. Patch orientations are optimized so that the required logical boundaries remain accessible, and the selected layout is then replayed with explicit routing-cell conflicts.

Candidate layouts are evaluated using the workload-dependent architecture objective. For each compiled resource candidate, the supplied layout backend returns a data/routing arrangement and a lattice-surgery (LS) schedule. Conflicting operations are serialized, and the resource evaluator then recomputes factory demand and checks error budgets. This interface preserves the separation between architecture synthesis and resource accounting while allowing candidate layouts to be compared using their complete modeled execution cost.

We compare Hybrid circuits on the standardized fast block and in a candidate pool that also permits circuit-specific layouts. Both arms use the same input circuits, physical error rate, factory catalogue, synthesis method, error budgets, LS scheduler, and reachable perimeter supply-interface convention. Code distance and provisioning may differ, so these are independently selected implementation plans rather than a frozen-resource spatial ablation. The optimized pool includes feasible fast candidates.

In the twenty-circuit comparison, the Hybrid layout search reduces runtime and volume in eighteen cases and retains the fast plan for QPE--factor and QPE--state preparation. Optimized-to-fast runtime ratios span $0.6912$--$1.0000$, and volume ratios span $0.6468$--$1.0000$. The median fast-to-optimized volume gain is $1.1215\times$. Including the fixed plan prevents a worse candidate from being selected; the improvements remain conditional on the finite search and declared supply model.

These results retain the architectural motivation for workload-dependent spatial organization without assuming that customization is universally preferable. The fast template and circuit-specific candidates provide alternative implementation choices; compact and intermediate blocks are not part of this common-scheduler comparison.

\subsection{Planning non-Clifford resources for fault-tolerant execution}

Non-Clifford resource preparation forms the second major resource-planning problem in the algorithm-to-execution workflow. Conventional surface-code implementations typically rely on dedicated magic-state factories that provide predictable throughput through continuously provisioned capacity. This strategy is effective when non-Clifford demand is sustained, but it can reserve substantial resources even when demand is sparse or strongly time dependent. The time-resolved non-Clifford demand retained in the executable workload therefore provides information that can be used to select how these resources are supplied during execution.

We treat the non-Clifford resource-generation strategy as part of the implementation rather than as a fixed backend assumption. We denote this choice by $\mathcal{F}$, which specifies the resource-generation mechanism, including factory configuration, cultivation parameters, and available production capacity. For a given executable workload $\mathcal{W}$, the corresponding implementation can be written as
\begin{equation}
\mathcal{F}^{*}
=
\arg\min_{\mathcal{F}}
C_{\mathrm{NC}}(\mathcal{F}|\mathcal{W}),
\label{eq:nonclifford_supply}
\end{equation}
where $C_{\mathrm{NC}}$ represents the physical cost of providing the required non-Clifford resources under the temporal execution demand of the computation. In the present framework, this choice is evaluated together with the other fault-tolerant execution parameters rather than optimized as an independent resource count.

Magic-state cultivation~\cite{gidney2024magicstatecultivationgrowing, claes2025cultivatingtstatessurface,chen2025efficientmagicstatecultivation, vaknin2026efficientmagicstatecultivation, sahay2026foldtransversalsurfacecodecultivation} provides a resource-state preparation protocol based on protected preparation and postselected verification; repeated cultivation units can themselves form a continuously provisioned factory fleet. It therefore introduces a different balance between persistent footprint, preparation latency, and acceptance probability. Recently, in-patch multiplexing~\cite{kim2026reducingpostselection} was proposed to reduce the postselection overhead of cultivation by placing multiple early-stage candidates within otherwise idle regions of a single logical patch. Resource-aware compilation has likewise begun to optimize placement, routing, scheduling, and the choice of magic-state protocol together~\cite{pflieger2026harvest}. The number and placement of simultaneous in-patch candidates remain constrained by the geometry of the predefined patch.

\begin{figure*}[htbp]
\centering
\includegraphics[page=1,pagebox=cropbox,width=\textwidth]{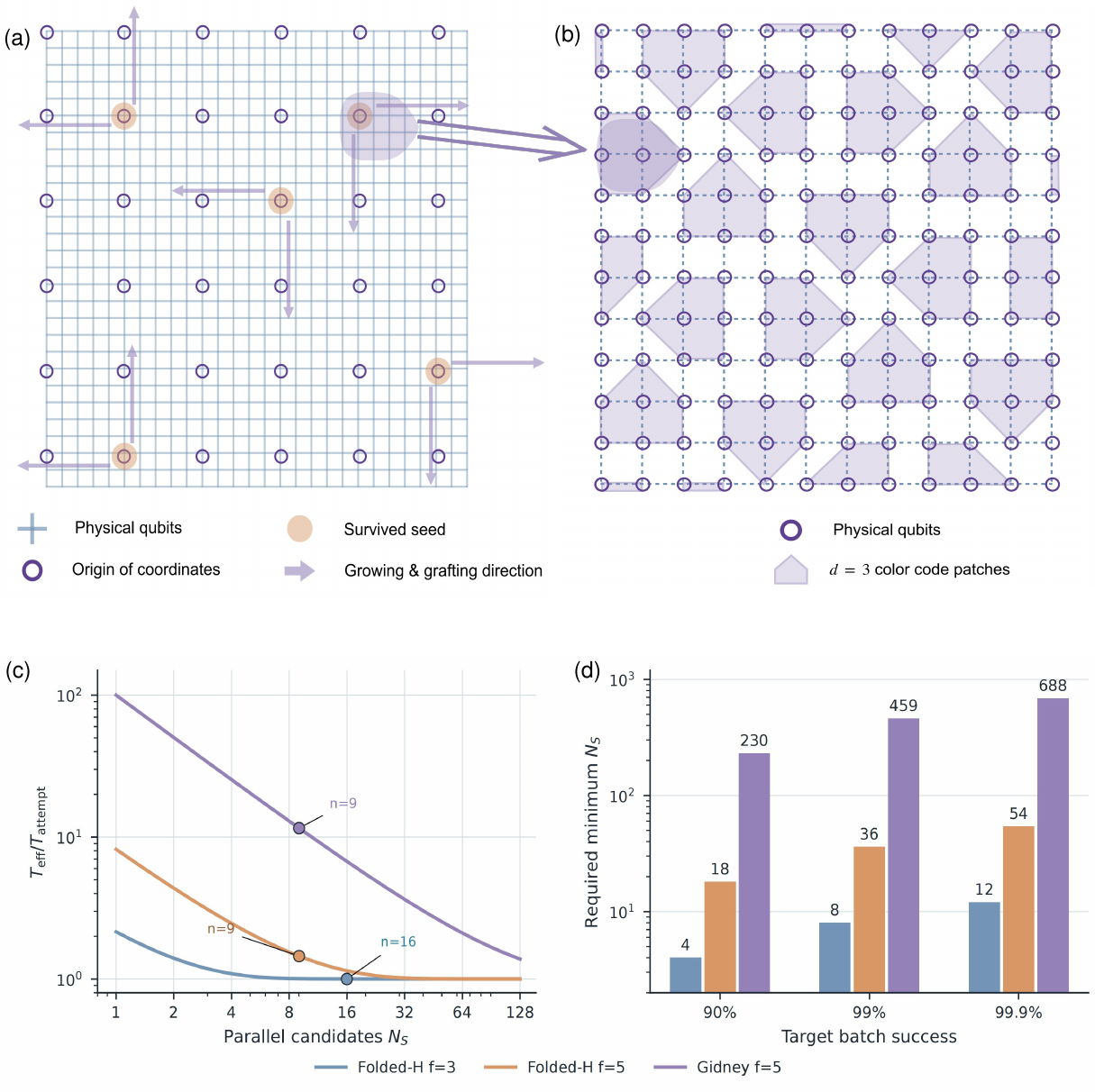}
\caption{\textbf{Pre-patch parallel cultivation as a resource-supply concept.} \textbf{(a,b)} Schematic seed placement and subsequent patch framing; these drawings illustrate the architectural proposal and are not simulated processor layouts. The arrow from (a) to (b) indicates that a coordinate origin in (a) represents an initial seed patch, which can be a distance-$3$ color-code patch of the form shown in (b). \textbf{(c)} The independent, fixed-duration batch relation $T_{\rm eff}/T_{\rm attempt}=1/(1-q^{N_S})$. Outlined dots mark the finite allocations used by the benchmark model: $N_S=16$ for Folded-H $f=3$, and $N_S=9$ for Folded-H $f=5$ and Gidney $f=5$. Their batch-success probabilities are approximately 0.999955, 0.689937, and 0.086483, respectively. \textbf{(d)} Ideal candidate counts required to reach selected batch-survival probabilities. Curves use catalogue discard rates 0.535, 0.878, and 0.990 in the same order. Large theoretical batches are not the benchmark allocations: the model uses $N_S=\lfloor D_2/D_1\rfloor^2$ and the whole-batch approximation in Supplementary Sec.~\ref{subsec:sm_ppc}. Correlated failures, geometric grafting feasibility, buffers, and queueing are outside this approximation.}
\label{fig:magic}
\end{figure*}

Here we introduce \emph{pre-patch parallel cultivation} (PPC), which extends in-patch multiplexing by placing cultivation candidates before the final logical patch boundary is assigned, as illustrated in Fig.~\ref{fig:magic}. Candidate injection sites are chosen before the final logical boundary is assigned, so the initial placement is not restricted to the geometry of one predefined patch. Each seed is represented by an initial patch location; after postselection, a surviving seed anchors the subsequent logical patch. The grow-and-graft direction and extent are then selected from locally available idle qubits, rather than being fixed before the survival outcomes are known. The architectural concept permits an available seeding region containing
\begin{equation}
N_S=\alpha N_Q
\end{equation}
spatially separated candidate sites, where $N_Q$ is the available physical-qubit count and $\alpha$ is an idealized seeding density. For the numerical benchmark model, the candidate count is approximated geometrically by $N_S=\lfloor D_2/D_1\rfloor^2$, using the selected protocol's small-code distance $D_1$ and final grafting distance $D_2$. These protocol distances are distinct from the data-code distance $d$. Once a candidate survives postselection, its location anchors the logical patch used for subsequent growth and grafting. Illustrative geometric constructions are given in the Supplemental Material.

If each candidate is discarded independently with probability $q$, the probability that at least one of the $N_S$ simultaneous candidates survives is
\begin{equation}
P_{\mathrm{survive}}(N_S)
=
1-q^{N_S}.
\label{eq:cultivation_survival}
\end{equation}
For a batch of fixed duration $T_{\mathrm{attempt}}$, repeating the batch until at least one candidate survives gives the effective preparation latency
\begin{equation}
T_{\mathrm{eff}}
=
\frac{T_{\mathrm{attempt}}}{P_{\mathrm{survive}}}
=
\frac{T_{\mathrm{attempt}}}{1-q^{N_S}},
\end{equation}
where $T_{\mathrm{eff}}$ is the mean total time to obtain the first surviving batch. The corresponding normalized latency is
\begin{equation}
\frac{T_{\mathrm{eff}}}{T_{\mathrm{attempt}}}
=
\frac{1}{1-q^{N_S}}
=
\frac{1}{1-q^{\alpha N_Q}}.
\label{eq:cultivation_latency}
\end{equation}
Figure~\ref{fig:magic}(c) evaluates this relation for three representative discard probabilities taken from the cultivation catalogue: Folded-H $f=3$ ($q=0.535$), Folded-H $f=5$ ($q=0.878$), and Gidney $f=5$ ($q=0.990$). These inputs span a broad range of rejection probabilities rather than exhaust the catalogue. Figure~\ref{fig:magic}(d) gives the minimum candidate counts required to reach selected batch-survival probabilities. The large idealized counts for Gidney $f=5$ illustrate that the required seeding area can exceed a realistic processor footprint; finite area, boundaries, seed separation, and grow-and-graft contention can therefore limit the achievable batch-survival probability. These curves isolate the postselection penalty by normalizing each protocol to its own attempt duration. They do not include correlated candidate failures, storage effects, or queueing.

The Gidney and Folded-H labels in Fig.~\ref{fig:magic}(c,d) identify the sources of the single-candidate discard probabilities. The plotted curves describe ideal independent batches, not the complete physical seed-farm implementation. Numerical estimates use each protocol's overall discard rate and finite $N_S$, retain its catalogue footprint, and rescale its expected output period by $(1-q)/(1-q^{N_S})$. They do not simulate intermediate growth failures, simultaneous grafting, or survivor transport. PPC is therefore an explicit throughput approximation in the benchmark resource model.

Within the independent-batch model, PPC exchanges parallel candidate capacity for lower expected preparation latency. The full processor-wide geometric benefit illustrated schematically has not been established by a physical growth-and-routing simulation. Its usefulness therefore depends on the execution context. Sustained high-throughput demand can favor continuously operating factories, whereas intermittent demand or temporarily available area can make cultivation-based preparation more attractive. The present results do not establish a universal boundary between these regimes, but they show that cultivation introduces a controllable trade-off between spatial footprint, postselection overhead, and preparation latency.

More generally, non-Clifford resource preparation need not be treated as a fixed backend component. Factory-based preparation, direct resource-state implementations, and cultivation provide different operating points in the execution design space, allowing the supply strategy to be selected according to the temporal demand and available physical resources of the scheduled computation.

\subsection{End-to-end execution reveals the physical cost of quantum algorithms}
\label{subsec:resource_landscape}

The preceding sections identify spatial organization and non-Clifford implementation as two major design choices in the algorithm-to-execution process. We now evaluate how these choices combine at the level of complete fault-tolerant executions. The resulting physical cost is not determined by logical circuit size alone, but by how the scheduled computation interacts with the selected surface-code layout, non-Clifford realization, and code distance.

We evaluate twenty concrete ZX/frame-transformed circuits spanning seven algorithm families. Logical compilation and architecture synthesis generate candidate circuit and spatial realizations, after which the resource evaluator selects among the resulting implementation choices according to
\begin{equation}
\widehat{x}
=
\arg\min_{x\in\mathcal C_{\rm evaluated}}
V_{\rm ST}(x),
\qquad
x=(\mathcal A,\mathcal F,d),
\label{eq:end_to_end_resource}
\end{equation}
subject to the surface-code and non-Clifford error budgets. Here $\mathcal A$ denotes the surface-code spatial organization, $\mathcal F$ the non-Clifford implementation and resource-supply configuration, and $d$ the code distance. The candidate set is finite and is generated through heuristic layout search and discrete resource sweeps, so Eq.~\eqref{eq:end_to_end_resource} identifies the best evaluated implementation rather than a certified global optimum.

To isolate the contribution of non-Clifford implementation, we compare two strategies. In All-$T$, every non-Clifford rotation is synthesized into a Clifford+$T$ sequence and is therefore realized entirely through $T$-state consumption. Hybrid enlarges this implementation space by allowing each eligible rotation to use either Clifford+$T$ synthesis or an available direct-rotation implementation under the same error and resource model. All-$T$ is thus contained in Hybrid as the special case in which all rotations are synthesized. Within Hybrid, the direct-rotation branch can further select among the admissible policy and provisioning options described above, while these internal choices are not treated as separate top-level strategies.

\begin{figure*}[htbp]
\centering
\includegraphics[width=0.85\textwidth]{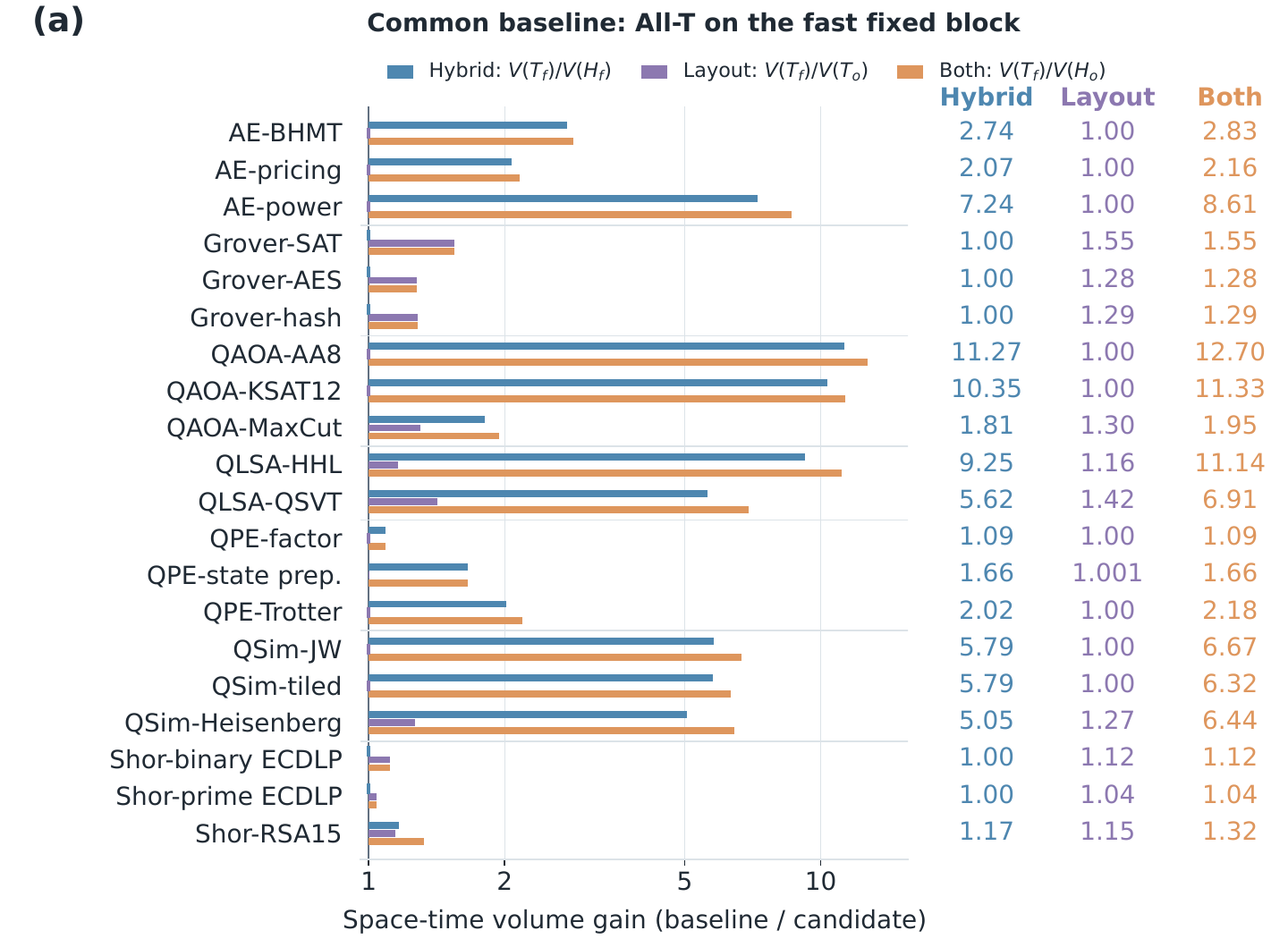}
\par\smallskip
\includegraphics[width=0.80\textwidth]{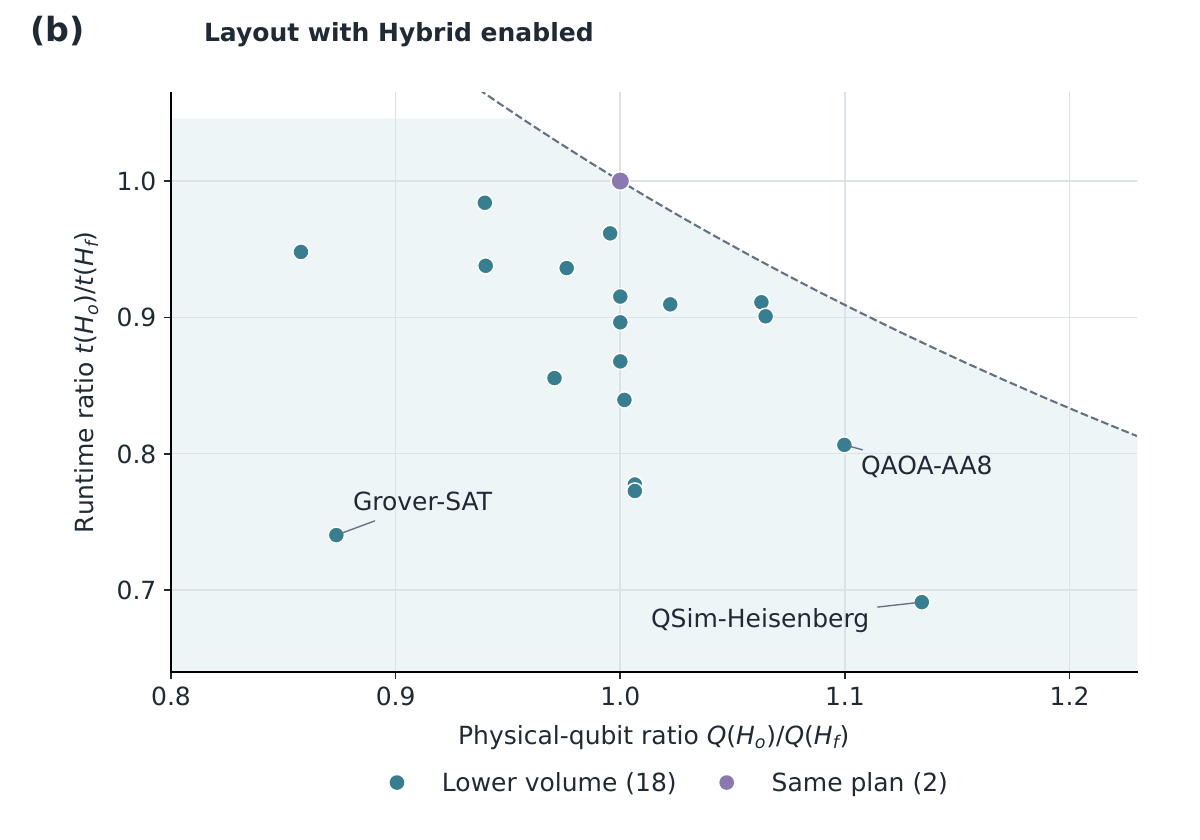}
\caption{\textbf{Hybrid and layout contributions across twenty circuits.} $T_f,H_f,T_o,H_o$ denote All-$T$/Hybrid on fast/optimized-pool layouts. \textbf{(a)} Common-baseline volume gains: Hybrid $V(T_f)/V(H_f)$, layout $V(T_f)/V(T_o)$, and both $V(T_f)/V(H_o)$; maxima are 11.27, 1.55, and 12.70. \textbf{(b)} Layout selection within Hybrid: qubit and runtime ratios relative to $H_f$. Teal points improve volume; purple denotes two retained fast plans. The dashed curve is equal volume. Arms independently select resources under the same corrected conditional mean-timing model.}
\label{fig:benchmark}
\end{figure*}

We separately vary the spatial organization. The fast fixed block of Ref.~\cite{Litinski2019gameofsurfacecodes} serves as the common baseline, while the optimized-layout pool contains this layout together with circuit-specific candidates generated by the architecture search. We denote the All-$T$ and Hybrid realizations on fast by $T_f$ and $H_f$, and their selected optimized-pool realizations by $T_o$ and $H_o$. Figure~\ref{fig:benchmark}(a) compares Hybrid alone, $V(T_f)/V(H_f)$; layout alone under All-$T$, $V(T_f)/V(T_o)$; and their combination, $V(T_f)/V(H_o)$. The supplementary tables additionally report the layout benefit within Hybrid, $V(H_f)/V(H_o)$, and the Hybrid benefit within the optimized pool, $V(T_o)/V(H_o)$. Figure~\ref{fig:benchmark}(b) resolves the space--time trade-off for $H_f\rightarrow H_o$. All ratios are baseline divided by candidate, and resource parameters are independently reselected for each arm. The optimized pool includes fast candidates, and Hybrid includes All-$T$, so each enlarged evaluated design space retains its corresponding baseline.

For the selected Hybrid optimized-layout implementations $H_o$, code distances range from $13$ to $19$, physical footprints from $8{,}135$ to $36{,}781$ qubits, and modeled runtimes from $1.989$ ms to $3.241$ s. The corresponding space--time volumes span $4.38\times10^7$--$2.38\times10^{11}$ physical-qubit--stabilizer-rounds under the $500$ ns stabilizer-round assumption, including external factory footprints.

We first isolate the effect of non-Clifford implementation on the fast block. Hybrid reduces space--time volume in fifteen of the twenty circuits, while the five exact-digital workloads give the same evaluated optimum as All-$T$. The Hybrid-only ratio $V(T_f)/V(H_f)$ reaches $11.2689\times$ for QAOA--AA8, with a median of $2.0434\times$. The synthesis-only realization remains admissible in every Hybrid comparison.

We next isolate spatial organization under All-$T$, comparing $T_f$ with $T_o$. Layout optimization lowers volume in eleven cases; $V(T_f)/V(T_o)$ reaches $1.5461\times$ for Grover--SAT, with a geometric mean of $1.1173\times$. With Hybrid already enabled, the separate comparison $H_f\rightarrow H_o$ lowers volume and runtime in eighteen cases, also reaching $1.5461\times$, with a geometric mean volume gain of $1.1437\times$. QPE--factor and QPE--state preparation retain fast. Figure~\ref{fig:benchmark}(b) shows that reduced runtime can compensate for a larger footprint.

Finally, comparing the All-$T$ fast baseline $T_f$ with Hybrid optimized $H_o$ improves all twenty circuits. The combined gain reaches $12.7048\times$ for QAOA--AA8, with a median of $2.1717\times$ and a geometric mean of $3.0470\times$. The combined gain decomposes as
\[
\frac{V(T_f)}{V(H_o)}
=
\frac{V(T_f)}{V(H_f)}
\frac{V(H_f)}{V(H_o)},
\]
separating the contribution of Hybrid non-Clifford implementation from that of spatial optimization.

The selected Hybrid plans further show how this gain is realized internally. Nineteen $H_o$ solutions use fixed-threshold policies; QPE--Trotter selects a one-trial fixed-RUS cap ($N=1$, $r=2$). Selection depends on the complete physical cost, including the analog history accumulated before digital fallback, factory demand, and direct-rotation area. R0 assigns one patch per $R_z$ slot and R1 three; after parameters are reselected, R0 has lower volume in fifteen circuits and ties R1 in the five cases with no direct rotations. These are supply-ready area scenarios rather than calibrated accepted-state throughput guarantees.

As an additional robustness check, six supplied parameter checkpoints per QAOA family are independently evaluated in all four arms. Combined gains span $12.7048$--$13.2592\times$ for AA8 and $13.3789$--$21.1770\times$ for KSAT12, with the checkpoint peak at KSAT12 train000. All-$T$ optimized retains fast at all twelve checkpoints, so its Hybrid ratios equal the combined ratios. Hybrid-on-fast gains span $10.0359$--$11.2689\times$ and $12.8025$--$18.7037\times$, respectively. These readout-aware instances are separate from the main twenty benchmarks and are not complete trajectories or shot-weighted training costs. Supplementary Table~\ref{tab:qaoa-trajectory} and Fig.~\ref{fig:sm_qaoa_trajectory} use the same corrected records.

The benchmark comparisons thus connect the logical workload, physical routing, and resource supply without assigning universal superiority to a single implementation. They are conditional execution estimates under the timing, error, and supply approximations stated in Methods and the Supplementary Information.

\section{Discussion}

The results of this work point to a broader conclusion: the physical cost of fault-tolerant quantum computation is not determined by logical resource counts alone, but emerges from the interaction between algorithm structure, fault-tolerant realization, and hardware organization. Circuits with similar logical widths or non-Clifford counts can lead to different physical costs once communication, spatial organization, resource-state supply, and execution timing are taken into account. Resource estimation at this level is therefore not only a counting problem, but also a cross-layer design problem.

The layout results illustrate this dependence directly. Circuit-specific spatial organization reduces space--time volume for eighteen of the twenty Hybrid benchmarks, while the remaining two retain the fast fixed layout. Because the fast layout is included in the candidate pool, these comparisons do not imply that every customized layout is advantageous. Instead, they identify the condition under which customization becomes useful: the reduction in communication and routing overhead must compensate for any additional spatial resources introduced by the layout. This suggests that the surface-code data block should not always be treated as a fixed compilation target, but can instead be organized according to the interaction structure and parallelism of the computation being executed. More broadly, future surface-code processors may benefit from treating logical patches as reconfigurable resources rather than fixed spatial objects. Code deformation and lattice surgery already provide fault-tolerant mechanisms for changing patch boundaries and connectivity~\cite{Horsman2012latticesurgery,Vuillot2019CodeDeformation}, while recent architectures have begun to exploit dynamic allocation of routing, ancilla, and resource-state regions~\cite{Kan2025sparo,tornow2026rushhour}. As hardware scales and physical error rates improve, this flexibility could extend from spatial placement to the protection itself, allowing logical patches to expand, contract, or change code distance across different phases of a computation. Such a dynamically reconfigurable QEC substrate would shift the optimization target from a static data block to the space--time allocation of logical protection and computational resources.

The same principle applies to non-Clifford implementation. Hybrid execution improves fifteen of the twenty benchmark circuits and reproduces the All-$T$ solution for the five exact-digital cases, showing that the preferred realization depends on the rotation structure of the workload rather than on a universally optimal synthesis rule. Direct rotations can avoid long Clifford+$T$ sequences, but their advantage depends on their error model, spatial footprint, fallback behavior, and the corresponding demand placed on $T$-state factories. Resource-state generation should therefore be considered together with the computation that consumes those states. The PPC model explored here provides one example of how temporary spatial resources may be exchanged for preparation latency and acceptance probability, but the present analysis does not establish a universal advantage for cultivation or a complete physical implementation of the associated seed-and-graft architecture.

The present framework coordinates algorithmic, architectural, and resource decisions across the execution stack, although the current implementation does not yet optimize all of these variables simultaneously within a single global search. Circuit-specific layouts are obtained from finite-budget heuristic searches and are subsequently evaluated together with non-Clifford resource choices and code distance. The execution model further approximates stochastic fallback by expected costs, represents non-Clifford demand through scheduled demand profiles rather than exact state-consumption timestamps, and does not fully embed all factory bodies, buffers, and delivery paths into a closed two-dimensional floorplan. Decoder effects are represented through an effective timing model, while the application-scale ECDLP example is reconstructed hierarchically rather than through a fully expanded physical schedule. These approximations limit the precision of individual numerical estimates, but do not alter the role of the framework as a common interface for exposing and comparing the coupled design choices that arise between logical algorithms and physical fault-tolerant execution.

A natural next step is to extend the present cross-layer optimization from static resource selection to time-dependent execution. Patch placement, lattice-surgery routing, resource-state generation and storage, congestion, and feedback dependencies all evolve during a computation and could therefore be coordinated within a common dynamic schedule. Stochastic models of resource-state production and fallback could further replace the present expected-demand approximations.

More broadly, the transition from logical demonstrations to useful fault-tolerant algorithms calls for a corresponding shift in how quantum-computing resources are analyzed. The relevant question is no longer only how many physical resources are required to execute a given logical circuit on a predefined architecture, but how the algorithm, fault-tolerant implementation, and hardware resources should be organized together to realize that computation efficiently. The algorithm-to-execution framework developed here provides one route toward this cross-layer view of fault-tolerant quantum computing.
\section{Methods}

\subsection{Algorithm representation and executable-workload construction}
\label{sec:algorithm-workload}

The framework starts from a quantum algorithm description together with its computational structure. Rather than reducing the computation directly to aggregate logical resource counts, we preserve its hierarchical organization, construct a dependency-preserving logical schedule, and retain the execution information required by the subsequent surface-code implementation. For explicitly generated circuits, this information is represented through an executable workload; for application-scale computations, the same quantities are reconstructed hierarchically from reusable computational modules.

\subsubsection{Hierarchical algorithm decomposition}

Quantum algorithms are represented hierarchically before fault-tolerant execution analysis. At the highest level, an algorithm is decomposed into reusable computational modules, such as arithmetic routines, oracle evaluations, simulation blocks, or phase-estimation procedures. Each module is recursively expanded into lower-level subroutines until reaching the elementary logical instruction set used by the fault-tolerant compiler.

\begin{figure*}[t]
\centering
\includegraphics[width=\textwidth]{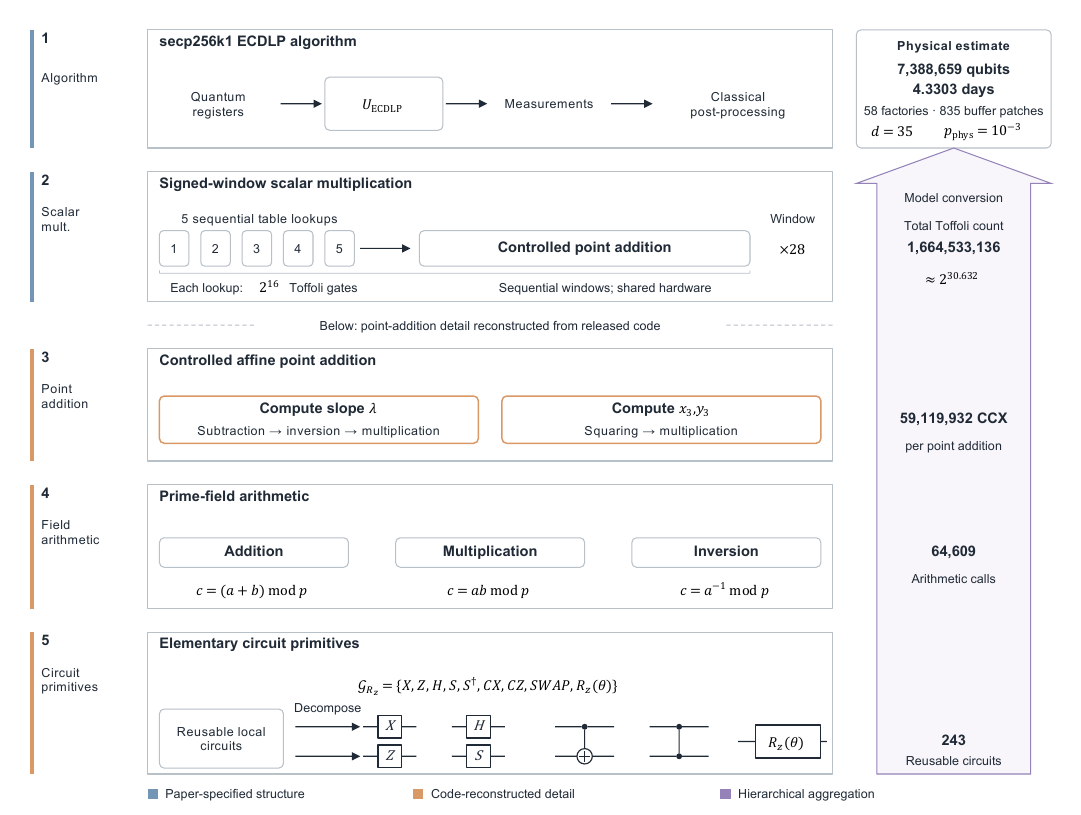}
\caption{\textbf{Hierarchical decomposition and bottom-up resource aggregation for the secp256k1 ECDLP computation.} Each of the 28 signed windows contains five table lookups and one controlled affine point addition, with a cost of $2^{16}$ Toffoli gates per lookup. The final measurement outcomes are converted into the discrete logarithm by classical post-processing. The point-addition and arithmetic decomposition below the dashed line is reconstructed from the released implementation, whereas the signed-window multiplicities above it are paper-derived. Upward arrows in the right column show the bottom-up aggregation of circuit-primitive resources into point-addition, algorithm-level Toffoli, and fault-tolerant physical-resource estimates. Composing the selected local LS profiles with a shared finite T-state buffer gives a hierarchical estimate of 7,388,659 physical qubits and 4.3303 days at code distance $d=35$, with 58 magic-state factories and 835 buffer patches. The buffer-search range and composition assumptions are specified in Supplementary Sec.~\ref{sec:sm_ecdlp}. This application-scale estimate is evaluated separately from the explicit-benchmark gain statistics.}
\label{fig:ecdsa_hierarchy}
\end{figure*}

For application-scale algorithms, explicitly expanding the complete elementary circuit is often impractical because the number of logical operations can become prohibitively large. Instead, the framework preserves the hierarchical structure of the computation and reconstructs execution resources through recursive composition of reusable modules. The application-scale secp256k1 computation is reconstructed hierarchically, as illustrated in Fig.~\ref{fig:ecdsa_hierarchy}. One modeled execution is organized from the paper-level signed-window scalar-multiplication and semiclassical-QFT structure down to controlled affine point addition, prime-field arithmetic, and reusable elementary circuit primitives. The decomposition below the dashed boundary is reconstructed from the released implementation, including the arithmetic and dynamic-event structure of the controlled point-addition module, whereas the signed-window multiplicities and semiclassical-QFT structure above the boundary follow the paper-level construction. Resource quantities are then aggregated in the reverse direction, from reusable local circuits to arithmetic subroutines, controlled point addition, the complete ECDLP workload, and finally the surface-code physical-resource estimate. This hierarchy allows the application-scale computation to be evaluated without explicitly materializing the complete elementary circuit while keeping the origin of each resource contribution traceable across abstraction levels.

Each lower-level module contributes elementary resources to the higher levels according to its invocation multiplicity. This recursive composition preserves the computational structure while enabling application-scale execution analysis without explicitly materializing the complete elementary circuit. The dynamic-event layer retains measurement, reset, classical-condition, and conditional-correction dependencies. All multiplicities below the controlled point-addition boundary are resolved from the implementation; at $n=256$, the EEA contribution is evaluated through the source-driven 1,476-step sequence.

If $N_g(v)$ denotes the number of elementary resources of type $g$ required by computational module $v$, the recursive composition is
\begin{equation}
N_g(v)
=
N_g^{\mathrm{intrinsic}}(v)
+
\sum_{e\in E(v)}
m_e N_g(c_e),
\label{eq:recursive_resource}
\end{equation}
where $E(v)$ denotes the set of child modules invoked by $v$, $c_e$ is the corresponding child module, and $m_e$ is its invocation multiplicity.

This hierarchical representation allows application-scale computations to be analyzed without expanding every elementary operation while preserving the information required for subsequent scheduling, surface-code organization, and physical resource evaluation. The recursive counts satisfy resource conservation at each validated aggregation boundary, with each parameter-dependent arithmetic family evaluated using its own concrete parameters.

\subsubsection{Logical circuit representation and scheduling}

For explicit instances, the logical compiler supplies an axis-aware circuit containing Pauli, Clifford, control, rotation, and measurement operations. The resource adapter accepts
\begin{equation}
\mathcal G=\{X,Y,Z,H,S^{\pm1},T^{\pm1},\mathrm{control},R_P(\varphi),M_P\},
\label{eq:rz-gate-set}
\end{equation}
with $P\in\{X,Z\}$ for rotations and $R_P(\varphi)=\exp(-i\varphi P/2)$. Control operations retain the two participating Pauli axes. The current twenty-circuit inputs are the supplied ZX/frame-transformed instances; this resource study does not replace the upstream logical-compilation algorithm.

Rotation angles are preserved until resource compilation. Exact digital angles are identified, other rotations are either synthesized or retained, and the resulting gate stream is rescheduled. H frames are propagated through axes and absorbed into terminal measurement bases when possible; quantum outputs requiring an actual H retain it. Original gate identifiers and available measurement metadata are carried in the handoff.

The logical circuit is converted into a dependency-preserving schedule. Let $Q(G_i)$ denote the qubit support of operation $G_i$, and let $G_i\prec G_j$ denote a directed precedence relation in the operation-dependency graph. Two operations can be assigned to the same logical layer only if
\begin{equation}
Q(G_i)\cap Q(G_j)=\varnothing,
\qquad
G_i\nprec G_j,
\qquad
G_j\nprec G_i.
\label{eq:disjoint-support}
\end{equation}
The explicit-benchmark resource adapter preserves operation order on each affected qubit and constructs earliest dependency layers from these edges. The supplied benchmark interface does not implement a general classical-condition scheduler. The separate hierarchical dynamic-event model retains measurement-to-condition and reset/reuse dependencies where these are available from the source reconstruction.

For each explicitly generated circuit, the resulting schedule provides the exact logical width, logical-layer count, operation parallelism, and time-resolved non-Clifford demand for the fixed circuit instance. For each layer $\ell$, we record the non-Clifford demand
\begin{equation}
r_{\mathrm{NC}}(\ell),
\end{equation}
which forms the temporal resource-demand profile used in subsequent implementation decisions. After a non-Clifford implementation strategy is selected, this demand is further resolved into the corresponding direct-rotation and $T/T^\dagger$ consumption traces.

For the application-scale ECDLP workload, each local circuit supplies an LS duration, timestamped T-state requests, a charged patch count, and quantum/classical port bindings. Within an arithmetic family, these profiles are scheduled in a deterministic dependency-preserving order subject to the fixed data-area constraint. Upper-level calls without complete bindings retain their declared serial order. The composed request trace is then evaluated with one shared factory fleet and a finite buffer whose inventory and production phase persist across module boundaries. Measurement, reset, and feed-forward dependencies are retained, with zero additional latency assigned to these events in the numerical model of Supplementary Sec.~\ref{sec:sm_ecdlp}.

\subsubsection{Executable-workload extraction}

From the dependency-preserving logical schedule, we extract the three execution descriptors used in the subsequent optimization. The weighted logical interaction graph is
\begin{equation}
G_{\mathrm{int}}=(V,E,W),
\end{equation}
where vertices represent logical qubits and the edge weights $W_{ij}$ characterize the frequency or time-weighted interaction demand between logical qubits $i$ and $j$.

Let $\ell=1,\ldots,L_{\mathrm{sched}}$ index the dependency-preserving logical layers. Operation parallelism is obtained from the schedule by recording the number of simultaneously executable operations in each logical layer,
\begin{equation}
P(\ell).
\end{equation}

The layer-dependent non-Clifford demand profile is
\begin{equation}
r_{\mathrm{NC}}(\ell),
\end{equation}
which records the scheduled demand for non-Clifford operations before the final implementation route is selected. Unlike a total non-Clifford count, this profile preserves the temporal variation of the demand.

Together, these quantities define the executable workload
\begin{equation}
\mathcal{W}
=
\left\{
G_{\mathrm{int}},
P(\ell),
r_{\mathrm{NC}}(\ell)
\right\},
\end{equation}
used as workload descriptors alongside the full ordered circuit. After resource-specific synthesis and frame propagation, the descriptors and logical layers are rebuilt rather than treating the original layer numbers as physical execution times.

\subsection{Surface-code execution model}
\label{sec:ft-execution-model}

We model fault-tolerant execution using a two-dimensional rotated surface-code architecture in which logical qubits are represented by surface-code patches. This model provides the common execution abstraction used to translate scheduled logical operations into lattice-surgery primitives, stabilizer-round timing, and physical resource requirements. It is not intended to reproduce the complete control stack of a specific hardware platform.

\subsubsection{Logical patches and fault-tolerant primitives}

Each logical qubit is encoded in a rotated surface-code patch with code distance $d$. The boundaries of a patch determine the available logical Pauli operators and the lattice-surgery operations that can be performed. Logical Pauli corrections are tracked using a classical Pauli frame rather than implemented through physical operations, so Pauli corrections contribute no additional quantum execution cycles in the resource model.

Non-Pauli logical operations are represented using fault-tolerant primitives compatible with two-dimensional surface-code architectures. Logical CNOT operations are implemented through lattice surgery, where ancilla regions mediate sequential logical $ZZ$ and $XX$ parity measurements between the participating patches, followed by Pauli-frame updates.

A physical logical Hadamard is represented through patch deformation and boundary transformation and costs three logical clocks. H operations absorbed into a classical frame or terminal measurement basis cost no clocks. Non-Pauli phase operations $S/S^\dagger$ cost one clock; Pauli operations cost zero. Physical boundary preparation and routing constraints are charged by the relevant scheduler.

Arbitrary logical $X$ or $Z$ rotations can be implemented through axis-appropriate resource-state teleportation. A prepared logical resource state
\begin{equation}
|m_\theta\rangle_L
=
R_{z,L}(\theta)|+\rangle_L
\end{equation}
is consumed through axis-appropriate gate teleportation. Pauli byproducts are tracked in the Pauli frame. For a non-Clifford angle, an unfavorable measurement outcome can also require a doubled-angle rotation correction; this correction is handled by the RUS process and its digital fallback described below~\cite{Toshio2026starmagicmutation}.

These primitives provide the execution interface between the scheduled logical computation and the physical surface-code model.

\subsubsection{Physical execution timing}

Following the usual lattice-surgery accounting convention, in which one logical time step roughly corresponds to $d$ code cycles \cite{Litinski2019gameofsurfacecodes}, we model one logical clock as $d$ syndrome-measurement rounds. Certain non-Clifford primitives and other operations whose basis choice or conditional branch depends on decoded information cannot proceed until the relevant correction is available. We assume that the protected wait for this result overlaps the $d$ stabilizer-extraction rounds performed during an otherwise idle logical clock. The effective duration of such a correction-dependent clock is therefore $\max(d\tau_{\mathrm{phys}},\tau_{\mathrm{rt}})$. Amortized over its $d$ rounds, the effective stabilizer-round duration is
\begin{equation}
\tau_{\mathrm{round}}
=
\max\left(
\tau_{\mathrm{phys}},
\frac{\tau_{\mathrm{rt}}}{d}
\right),
\label{eq:round-time}
\end{equation}
where $\tau_{\mathrm{phys}}$ is the physical syndrome-extraction time per round and $\tau_{\mathrm{rt}}$ is measured from the availability of the last required syndrome data to delivery of the decoded correction. Thus, $\tau_{\mathrm{rt}}/d$ is an amortized contribution to the correction-dependent logical-clock model.

We decompose the real-time feedback latency as
\begin{equation}
\tau_{\mathrm{rt}}
=
\tau_{\mathrm{in}}
+
\tau_{\mathrm{decode}}
+
\tau_{\mathrm{out}},
\label{eq:decoder-feedback}
\end{equation}
where $\tau_{\mathrm{in}}$ includes syndrome aggregation and transport to the decoder, $\tau_{\mathrm{decode}}$ is the decoder response time after the required data become available, and $\tau_{\mathrm{out}}$ is the time required to deliver decoded information to the runtime controller.

Full-system decoding demonstrations provide reference values for different parts of this feedback path. A three-RFSoC RISC-Q prototype reported $446$ ns from availability of the final syndrome to return of feedback for an emulated distance-3 workload, with a $56$ ns Helios decoder core \cite{Liu2026riscq}. On Rigetti's Ankaa-2 processor, an eight-qubit stability experiment measured a $9.6\,\mu$s response from the final readout to a conditional gate for a nine-round batch, comprising $6.5\,\mu$s of decoding and $3.1\,\mu$s of communication and control \cite{Caune2024realtime}. Willow's distance-5 streaming decoder reported $63\pm17\,\mu$s from receipt of the final syndrome cycle to production of a correction while sustaining $1.1\,\mu$s QEC cycles; input transmission was estimated to take less than $10\,\mu$s, and output feedback was not implemented \cite{Acharya2024belowthreshold}.

Full-system measurements remain scarce and cover only a limited range of platforms and code distances. Hardware and software decoder benchmarks therefore provide broader, although less complete, constraints on decoder-side timing. Sparse Blossom and Helios report finite-block runtimes amortized per syndrome round, whereas Micro Blossom measures the response from syndrome availability to correction-bit availability, including CPU--FPGA I/O \cite{HiggottGidney2023sparseblossom,Wu2025microblossom,Liyanage2024helios}. Collision Clustering reports decoder-core times for FPGA implementations and a post-layout ASIC design rather than a measured end-to-end control path \cite{Barber2023collisionclustering}. Belief-matching and belief-find improve circuit-level decoding accuracy, while parallel-in-time windowing addresses throughput and syndrome backlog \cite{Higgott2022beliefmatching,Tan2022parallelTime}. AQ2-RT sustains sub-$1\,\mu$s-per-cycle throughput only up to $d=11$, whereas a self-coordinating decoder reports $0.979\pm0.021\,\mu$s per round at $d=25$ on a single TPU v6e using simulated long-memory workloads \cite{Senior2025alphaqubit2,Zhang2026selfcoordinatingNN}. These results constrain decoder selection and accelerator provisioning but do not determine a unique end-to-end $\tau_{\mathrm{rt}}$.

We therefore retain decoder timing as an explicit execution parameter whose effect depends on when decoded information is actually required. Corrections that can be commuted into the Pauli frame are removed from the critical execution path, whereas the feedback term in Eq.~\eqref{eq:round-time} applies when a subsequent basis choice or branch depends on decoded information. This synchronization semantics is consistent with the three-layer architecture of Ref.~\cite{Wang2026threeLayerFTQC}, in which a configured decoder consumes syndrome measurements and sends correction or Pauli-frame information to a controller that waits only at correction-dependent boundaries. For such a logical clock,
\begin{equation}
\tau_{\mathrm{clock}}
=
d\tau_{\mathrm{round}}.
\label{eq:logical-clock}
\end{equation}
Clocks that do not depend on decoder output retain a duration of $d\tau_{\mathrm{phys}}$ under this model.

Although the analytical model retains $\tau_{\mathrm{round}}$ as a free parameter, a numerical resource estimate requires a concrete reference scenario. We therefore use $\tau_{\mathrm{round}}=500$ ns for the baseline calculations. In this numerical scenario we do not evaluate Eq.~\eqref{eq:round-time} from separate physical-cycle and decoder-latency inputs; 500 ns is the assumed effective round duration, not a measured decoder-limited value. State-preparation startup is excluded under steady-state pre-supply.

\subsubsection{Non-Clifford resource interfaces}

Universal fault-tolerant computation requires non-stabilizer resources in addition to Clifford operations. In the execution model, these resources are treated as components whose generation and consumption are determined after the scheduled non-Clifford demand has been extracted.

A non-Clifford rotation can follow two implementation paths. Rotations assigned to digital implementation are decomposed into Clifford+$T$ operations and consume prepared $T$ states. Alternatively, selected small-angle rotations can be implemented through direct logical resource-state preparation and teleportation.

The execution model therefore retains rotation angles until the non-Clifford implementation stage, where each rotation can be assigned to synthesis or direct resource-state implementation under the physical model described below.

\subsubsection{Physical resource accounting}

The processor is represented as a collection of logical patches allocated to data storage, routing and ancillary operations, direct-rotation resources, and interfaces to non-Clifford resource generation. For a selected implementation, the physical cost is determined from the number of grid-resident logical patches, the code distance, the execution duration, and the resource-state supply requirements.

This execution abstraction provides the common physical model used by the circuit-specific architecture synthesis, non-Clifford implementation, and end-to-end evaluation described below.

\subsection{Circuit-specific surface-code architecture synthesis}
\label{sec:architecture-synthesis}

The executable workload provides the interaction and resource-demand information used to construct a circuit-specific surface-code layout. We represent the processor as a two-dimensional grid of logical tiles assigned to data patches, routing and ancillary regions, and non-Clifford resource interfaces. The objective is to reduce routed execution overhead while controlling the additional spatial footprint introduced by customization.

\subsubsection{Surface-code layout representation}

A candidate architecture is represented by an assignment of logical qubits to positions on a two-dimensional surface-code grid together with the boundary orientation of each logical patch. Each data patch is characterized by its available $X$ and $Z$ boundaries, which determine the lattice-surgery operations that can be performed with neighboring routing or ancillary regions.

The architecture must support several classes of fault-tolerant operations. Two-qubit interactions require valid routing paths between the corresponding logical patches. Non-Clifford operations require access to resource-state interfaces. Basis-changing Clifford operations impose compatibility constraints on logical patch boundaries. The resulting layout therefore depends on the scheduled interaction and resource-access requirements of the computation.

\subsection{Implementation details of the layout objective}

The circuit-specific layout search uses the static objective defined in the main Methods, which combines interaction-weighted communication distance, non-Clifford resource access, and logical-boundary compatibility. Here we specify how these quantities are evaluated on the discrete surface-code grid.

For a required two-patch operation between logical qubits $i$ and $j$, the distance $d(i,j)$ is defined as the length of the shortest path between compatible logical boundaries through routing cells that are not occupied by data patches. The boundary orientation of each patch is included in the search, so two layouts with identical qubit positions but different boundary assignments can have different routing costs.

For non-Clifford access, $d(i)$ denotes a distance to a resource-access boundary in the architecture abstraction. The resource interface returns a data/routing layout and an Rz slot capacity; complete factory bodies and internal factory transport are not embedded by this interface. Their catalogue footprints are added separately. A visible access site must not be interpreted as a full factory floorplan.

Boundary-access penalties are evaluated from the patch orientation required by the assumed logical primitive. A candidate layout is rejected when a required boundary configuration cannot be supported by the surrounding routing geometry. The architecture-side proxy is distinct from the resource objective. The benchmark wrapper uses the supplied backend and evaluates its returned candidates without modifying its placement or routing algorithm.

The static objective is used only during layout search. The selected geometry is subsequently evaluated by replaying the original logical schedule with explicit routing-cell conflicts, so the search objective and the reported routed execution time are not identical quantities.
\subsubsection{Layout optimization and routed evaluation}

The architecture-synthesis problem is a discrete optimization over logical-qubit placement and patch orientation. Because the search space grows rapidly with circuit width, we use a heuristic genetic optimization procedure.

Each candidate represents a valid assignment of logical qubits and patch orientations on the surface-code grid. New candidates are produced through crossover and mutation. Crossover combines two parent layouts by transferring selected logical-qubit assignments while preserving one-to-one placement. Mutation introduces local changes by exchanging patch assignments or modifying boundary orientations.

For the resource experiments, rectangular interiors with area from the logical width $n$ to twice the fast-template area are screened in both orientations, together with previously used dimensions. A short operation-coverage circuit is evaluated at seed 11, five iterations, and initial/population sizes 8/8. Several shapes are retained for full-circuit evaluation. Up to two compiled structures per policy family and two geometries per structure receive refinement at seeds $(11,29,47)$, 50 iterations, and sizes 20/20. This is a finite heuristic search rather than a global optimality certificate; details are in the Supplementary Information. The reported records incorporate the complete reachable analog-error history and the RP$^2$ $f=3$ footprint correction. Distance, grouping, and factory provisioning are reselected over the retained verified policy/layout winners from the preceding search and its provisioning branches; the correction stage does not constitute a new exhaustive geometry search.

Both the literal fast template and circuit-specific candidates are evaluated with the same supplied LS scheduler and reachable perimeter T/Rz interfaces. Shared two-qubit fast patches are represented explicitly; an unused half for odd width remains charged. A resource-side legalizer separates conflicting operations while preserving dependencies. If a direct control remains rejected, it may use an exact four-control decomposition through an existing data mediator, whose state is restored; all extra LS operations and dependencies are charged. A candidate is rejected if no verified realization is found. Each selected lowered circuit is replayed on a freshly constructed layout. Compact/intermediate scheduling and the backend's gate-depth fallback are not used in the reported comparison.

The architecture backend proposes candidates using its layout-side objective. The outer resource evaluator ranks feasible returned candidates by physical space--time volume after recomputing supply and error-correction parameters. This coupling does not change the underlying architecture-synthesis method.

\subsection{Non-Clifford implementation and resource supply}
\label{sec:nonclifford-estimation}

Non-Clifford operations introduce both an implementation choice and a resource-supply requirement. Rotations can be assigned to Clifford+$T$ synthesis or direct resource-state implementation, after which the resulting $T/T^\dagger$ demand determines the required magic-state supply. These choices are evaluated using the scheduled non-Clifford demand retained in the logical representation.

\subsubsection{Hybrid rotation implementation}

For $R_P(\varphi)=\exp(-i\varphi P/2)$, write $\varphi=m\pi/2+\beta$ with $|\beta|\leq\pi/4$ and retain the Clifford factor. The residual magnitude is
\begin{equation}
a(\varphi)=|\varphi-(\pi/2)\operatorname{round}(2\varphi/\pi)|.
\label{eq:angle-reduction}
\end{equation}
Exact multiples of $\pi/4$ are implemented digitally without approximation. Other digital rotations are synthesized into actual Clifford+$T$ words using Ross--Selinger synthesis~\cite{RossSelinger2014optimal}. Each word is checked, including its scalar phase, against the target in operator norm,
\begin{equation}
\|e^{i\gamma}U_{\rm word}-R_P(\varphi)\|\leq\epsilon_{\rm syn}.
\label{eq:T-synthesis}
\end{equation}
Gate counts and timings come from these words, not a logarithmic T-count proxy.

The fixed-threshold branch retains non-exact residuals with $a(\varphi)<\theta_{\rm th}$ and doubles the feedback angle until the cutoff or an exact digital angle is reached. Threshold candidates include the actual angle-induced switching boundaries. A parallel branch caps the number of analog trials at $N=1,\ldots,8$ (nominal ratio $r=2^N$), with earlier termination at the validity limit of the TMR angle model. For effective length $L_i$, the fallback probability and mean analog trial count are
\begin{equation}
p_i=2^{-L_i},\qquad \bar n_i=\sum_{j=0}^{L_i-1}2^{-j}=2(1-p_i).
\end{equation}
The full conditional digital correction is synthesized with the same method as deterministic digital rotations. Its mean-work duration is included as
\begin{equation}
\bar C_i=\bar n_i+1+p_i C_{{\rm fb},i},
\end{equation}
where $C_{{\rm fb},i}$ includes the surviving Clifford operations, T injections, and required boundary operations. This is an execution-time approximation under pre-supply, not a claim about state-production throughput.

Analog errors are evaluated from the angle-specific leading-order STAR/TMR model~\cite{Toshio2026starmagicmutation}, using the paper angle $\theta_{L,i}=|\beta_i|/2$. The main hardware convention is $P_{\rm ud}=(k/15)p_{\rm phys}$, with $2\leq k\leq\lfloor d/2\rfloor$. The effective $\alpha_{{\rm RUS},i}=P_{{\rm analog},i}/(\theta_{L,i}p_{\rm phys})$ is computed for each angle and stopping rule; it is not fixed to one. The equations, coefficient assumptions, and supply limitations are given in Supplementary Sec.~\ref{sec:sm_nonclifford}.

\subsubsection{Magic-state generation and supply}

Rotations assigned to Clifford+$T$ synthesis consume prepared $T/T^\dagger$ states. Candidate factory protocols are selected according to both required output fidelity and scheduled production demand.

All original, exactly digitized, synthesized, and fallback T operations use the same selected factory type. With expected demand $\bar N_T$, the accepted-state output error must satisfy
\begin{equation}
p_T^{\rm out}\leq p_T^{\rm tar}
=\frac{\epsilon_{\rm NC}-P_{\rm analog}-P_{\rm syn}}{\bar N_T},
\label{eq:T-target-error}
\end{equation}
when $\bar N_T>0$ and the residual budget is nonnegative. No factory is required when the demand is zero.

For distillation-based factories, the output fidelity follows the corresponding protocol-dependent logical-error model. For example, an ideal $15$-to-$1$ Bravyi--Kitaev distillation block satisfies
\begin{equation}
p_{\mathrm{out}}
=
35p^3+O(p^4).
\end{equation}

For postselected preparation protocols, we distinguish the residual error probability of an accepted state from the probability that an attempted preparation is rejected. The accepted-state residual error contributes to $\epsilon_{\mathrm{NC}}$, whereas rejection is heralded and contributes to preparation latency and throughput rather than directly to the logical-error budget.

If a cultivation attempt is accepted with probability
\begin{equation}
p_{\mathrm{succ}}=1-q,
\end{equation}
where $q$ is the discard probability, its expected production overhead is
\begin{equation}
C_{\mathrm{cult}}
=
\frac{C_{\mathrm{attempt}}}
{p_{\mathrm{succ}}}.
\label{eq:cultivation-cost}
\end{equation}
For pre-patch parallel cultivation (PPC), simultaneous candidates instead give a batch-survival probability $1-q^{N_S}$ as described in the Results. The corresponding rejection probability affects expected preparation latency and resource availability but is not counted as an undetected logical error.

After scheduling each candidate, the backend's per-layer T counts generate unit-weight request events at cumulative layer starts, and fallback T events carry their branch probabilities. These timestamps are an explicit approximation, not verified state-consumption times. For factory output period $c_F$, the continuous-window provisioning rule is
\begin{equation}
N_F=\left\lceil\max_t\sum_{a:\,t\leq t_a<t+c_F}w_a\right\rceil.
\label{eq:factory-number}
\end{equation}
The period is adjusted by PPC as $c_F=c_{F,\rm cat}(1-q)/(1-q^{N_S})$ when the catalogue already includes single-candidate repetition. Multiplicity is recomputed after layout scheduling, and this expected-demand rule assumes pre-supply. It is used for the explicit benchmarks. The ECDLP calculation instead evaluates the deterministic request trace with continuous finite inventory and jointly selects the factory count and buffer capacity, as described in Supplementary Sec.~\ref{sec:sm_ecdlp}.

Direct-rotation supply is treated identically for fast and circuit-specific layouts. The LS evaluator limits concurrent retained rotations by the provided Rz slots; R0 charges one patch per slot and R1 three. These are alternative area assumptions over the same verified schedule pool, with resource parameters reselected, rather than a calibrated comparison of accepted-state rates. No additional fixed-block three-clock demand window is applied. Finite-noise postselection has not been fed back into supply; details are in the Supplementary Information.

\subsection{Application-scale algorithm composition}

For application-scale computations whose complete elementary circuits cannot be materialized explicitly, we apply the same algorithm-to-execution workflow hierarchically. Local circuit structure is retained where explicit bindings and schedules are available, while higher-level resources are composed through validated module multiplicities and conservative scheduling assumptions.

For the 256-bit secp256k1 elliptic-curve discrete-logarithm computation considered here, the strongest source-reconstructed resource-counting boundary is one controlled affine point addition, corresponding to Fig.~14 of Luo~et~al.~\cite{luo2026quantumalgorithmellipticcurve}. This subroutine has a peak quantum width of 835 logical qubits and contains $59{,}119{,}932$ CCX gates, $99{,}644{,}941$ CX gates, $24{,}105{,}455$ measurements, and $24{,}105{,}455$ resets. Measurement-conditioned corrections and subsequent qubit reuse are retained as dynamic dependencies.

The arithmetic leaf frontier contains 64,609 subroutine invocations drawn from 1,931 unique parameter families. For layout handoff, these families admit exact ordered compositions over 243 reusable exact local circuits, including 220 two-to-six-qubit circuits and 23 single-qubit or dynamic-event circuits. Their gate and dynamic-event resources are composed according to the algorithmic hierarchy rather than expanded into one explicit gate-level circuit.

At the paper-derived top level, the controlled point-addition resource is combined with the signed-window and table-lookup multiplicities of the original construction. For $n=256$ and window size $w=16$, one execution contains 28 controlled point additions and five sequential table lookups per window, for a total of 140 lookups. The resulting Toffoli count is
\begin{equation}
\begin{aligned}
N_{\mathrm{CCX}}^{\mathrm{full}}
&=
28\left(59{,}119{,}932+5\times2^{16}\right)\\
&=
1{,}664{,}533{,}136\\
&=
2^{30.63247044}.
\end{aligned}
\label{eq:luo-full-toffoli}
\end{equation}
The controlled point additions contribute approximately $99.45\%$ of this modeled Toffoli count, while table lookups contribute approximately $0.55\%$.

Using a fixed seven-$T$ decomposition for each CCX, the Toffoli contribution to the non-Clifford demand is
\begin{equation}
N_T^{\mathrm{CCX,full}}
=
7N_{\mathrm{CCX}}^{\mathrm{full}}
=
11{,}651{,}731{,}952.
\label{eq:luo-full-t-count}
\end{equation}
The paper-derived semiclassical-QFT structure contains 513 nontrivial rotations. For this part of the hierarchical estimate, we retain a fixed synthesis cost of 67 $T$ states per rotation, adding
\begin{equation}
513\times67=34{,}371
\end{equation}
$T$ or $T^\dagger$ states. The modeled one-run demand is therefore $11{,}651{,}766{,}323$ consumed $T/T^\dagger$ states. The QFT contribution retains its fixed timing and synthesis-error allocation, separately from the explicit Ross--Selinger words used for the benchmark circuits.

We compose the selected local LS profiles through the arithmetic hierarchy while retaining the paper-derived serial boundaries between upper-level modules and table lookups. The resulting base duration is $20{,}926{,}210{,}682$ logical clocks. Global T-state supply adds $453{,}292{,}158$ waiting clocks under the selected factory and buffer configuration, giving
\begin{equation}
C_{\rm exec}=C_{\rm LS}+C_{\rm wait}=21{,}379{,}502{,}840.
\label{eq:luo-full-execution-clocks}
\end{equation}
The largest individual request contains four T states. The QFT and readout contribution retains the prescribed 514-round semiclassical structure and its fixed timing model.

Under the compact-block physical-resource convention used for this reconstruction, a compact data block for $N_{\mathrm{logical}}$ algorithm qubits contains
\begin{equation}
\left\lceil1.5N_{\mathrm{logical}}\right\rceil+3
\end{equation}
data patches and assigns $2d^2-1$ physical qubits to each grid-resident patch. Thus, the peak width of 835 logical qubits requires
\begin{equation}
\left\lceil1.5\times835\right\rceil+3
=
1{,}256
\end{equation}
data patches. At $p_{\mathrm{phys}}=10^{-3}$, the lowest-volume evaluated configuration uses code distance $d=35$, 58 instances of the $(15\text{-to-}1)^6\times(15\text{-to-}1)$ factory, and a buffer of $B=835$ T-state patches. Each factory has a 39,100-qubit footprint, output error $3.3\times10^{-14}$, and output cadence 97.5 logical clocks. Buffer patches remain provisioned throughout execution. The selected buffer lies at the lower end of the declared search range in Supplementary Sec.~\ref{sec:sm_ecdlp}.

The data patches require
\begin{equation}
N_{\mathrm{data}}
=
1{,}256(2\times35^2-1)
=
3{,}075{,}944
\end{equation}
physical qubits. The buffer requires $835\times2{,}449=2{,}044{,}915$ physical qubits, and the selected factories require
\begin{equation}
58\times39{,}100
=
2{,}267{,}800
\end{equation}
physical qubits, giving a peak modeled footprint
\begin{equation}
\begin{aligned}
N_{\mathrm{phys}}
&=(1{,}256+B)(2d^2-1)+N_FQ_F\\
&=3{,}075{,}944+2{,}044{,}915+2{,}267{,}800\\
&=7{,}388{,}659.
\end{aligned}
\label{eq:luo-full-physical-qubits}
\end{equation}
With a 500-ns stabilizer round, the estimated runtime is $374{,}141.299700$ seconds, or approximately 103.9281 hours (4.3303 days). The corresponding space--time volume is $5.528804963\times10^{18}$ physical-qubit--stabilizer-rounds. The data block, buffer, and factory fleet are shared across all 28 windows, with inventory carried continuously between them.

This application-scale result describes one execution under the paper-derived top-level composition, fixed compact data area, and local LS timing model. The source-level point-addition resources and paper-level multiplicities are combined with a common global supply configuration. 

\subsection{End-to-end physical resource evaluation}
\label{sec:end-to-end-estimation}

The final stage combines the logical and architectural handoffs with resource and error-correction parameters. For explicit benchmarks, an outer candidate search couples the rotation policy, TMR grouping, returned layout and schedule, Rz capacity, factory type/multiplicity, and code distance. The architecture backend remains a separate component; the evaluator selects the lowest modeled volume among feasible evaluated candidates. The application-scale ECDLP reconstruction uses its separate hierarchical model.

\subsubsection{Error budget and code distance}

We allocate separate surface-code and non-Clifford error-proxy budgets,
\begin{equation}
\epsilon_{\mathrm{tot}}
=
\epsilon_{\mathrm{SC}}
+
\epsilon_{\mathrm{NC}},
\qquad
\epsilon_{\mathrm{SC}}
=
\epsilon_{\mathrm{NC}}
=
10^{-2}.
\end{equation}
The non-Clifford allocation includes residual magic-state errors, direct-rotation errors, and synthesis errors, giving the additive error-proxy ceiling
\begin{equation}
\epsilon_{\mathrm{tot}}
\leq
2\times10^{-2}
\end{equation}
used in the reported estimates. Postselection rejection in distillation or cultivation is not included in $\epsilon_{\mathrm{NC}}$ because it is heralded; it contributes instead to expected preparation cost, latency, and throughput. Residual errors in accepted resource states do contribute to $\epsilon_{\mathrm{NC}}$.

For physical error rate $p_{\mathrm{phys}}$, the logical error rate of a surface-code patch is modeled as
\begin{equation}
p_L(p_{\mathrm{phys}},d)
=
A
\left(
\frac{p_{\mathrm{phys}}}
{p_{\mathrm{th}}}
\right)^{(d+1)/2},
\label{eq:surface-logical-error}
\end{equation}
with $A=0.1$ and $p_{\mathrm{th}}=10^{-2}$. The numerical baseline uses $p_{\mathrm{phys}}=10^{-3}$.

For the explicit benchmarks, let $C_{\rm exec}=\lceil\bar C_{\rm sched}\rceil$ denote the charged logical-clock duration after physical scheduling, and define the protected-patch proxy
\begin{equation}
L_{\rm proxy}=N_{\rm grid}+N_F L_F.
\end{equation}
Here $N_{\rm grid}$ comprises charged data, routing, and direct-rotation patches; perimeter access markers are not additional charged patches. $L_F$ is the catalogue logical-patch count per factory. Then
\begin{equation}
P_{\rm SC}\simeq L_{\rm proxy}C_{\rm exec}d\,p_L(p_{\rm phys},d)
\leq\epsilon_{\rm SC}.
\label{eq:surface-error-budget}
\end{equation}
The accepted-state factory error is counted separately in $\epsilon_{\rm NC}$; the catalogue factory-patch exposure remains in $P_{\rm SC}$ by the declared accounting convention. Odd distances $5\leq d\leq37$ are searched together with all compatible $k$, factory types, and candidate schedules. The chosen distance minimizes feasible volume, rather than being fixed before resource selection. For ECDLP, the search uses $d=5,7,\ldots,37$ and includes buffer exposure through $L_{\rm proxy}=1{,}256+B+N_FL_F$; execution time includes the global supply waits. Its surface-error screening threshold is 0.009, with final verification against the 0.01 budget.

The computation-level error budget is fixed across workloads; the per-round logical error rate is not. Longer or larger computations therefore require a smaller admissible $p_L$ and may select a larger code distance. Reliability overhead is consequently absorbed primarily into the fault-tolerant implementation through code distance and physical resource cost rather than through an unconstrained repetition factor.

\subsubsection{Physical resource calculation}

The explicit benchmarks use the following patch-footprint convention; the independent ECDLP physical reconstruction is specified separately in Supplementary Sec.~\ref{sec:sm_ecdlp}. Each grid-resident data, routing/ancilla, or direct-rotation patch contributes
\begin{equation}
Q_{\mathrm{patch}}
=
2d^2-1
\end{equation}
physical qubits, while factory catalogue footprints are absolute physical-qubit counts. Thus,
\begin{equation}
N_{\mathrm{phys}}
=
N_{\mathrm{grid}}(2d^2-1)
+
N_FQ_F.
\label{eq:physical-qubits}
\end{equation}
Here $N_{\rm grid}=N_{\rm data+route}+N_R$ for both layout families, including the unused fast half-tile when present. Perimeter T/Rz markers describe supply access, not separately charged factory bodies or extra grid patches; the Rz fleet and external factory footprint are counted explicitly. Factory footprints are absolute physical-qubit counts and are not multiplied by $2d^2-1$.

For explicit benchmarks, runtime is obtained from the routed, duration-weighted schedule, not the original logical layer count:
\begin{equation}
T_{\rm exec}=C_{\rm exec}d\tau_{\rm round},
\label{eq:runtime}
\end{equation}
with $\tau_{\rm round}=500$ ns. In both arms, residual physical H corrections on distinct quantum outputs are charged as a terminal three-clock layer. Terminal measurements remain in the logical handoff but are omitted from the LS timing trace; no independent readout latency is included. The same endpoint convention is used in all comparisons.

The provisioned volume is
\begin{equation}
V_{\rm ST}=N_{\rm phys}C_{\rm exec}d,
\label{eq:spacetime}
\end{equation}
in physical-qubit--stabilizer-rounds; multiplication by $\tau_{\rm round}$ gives qubit-seconds. Peak provisioned footprint is charged for the full modeled duration. Candidate ratios satisfy $G_V=G_QG_T$. Full search settings, assumptions, and circuit-resolved values are given in the existing Supplementary sections.

\subsubsection{Hierarchical physical-resource evaluation}

For computations whose elementary circuits cannot be explicitly expanded, the end-to-end evaluator operates on the hierarchical representation introduced above. Module-level resource profiles are composed according to invocation multiplicity. Additive resources are summed, while peak logical width is determined from the declared register structure and reuse assumptions rather than by summing child widths.

For modules $M_a$ with repetition counts $n_a$, the total non-Clifford demand is accumulated as
\begin{equation}
N_T
=
\sum_a
n_aN_T(M_a),
\end{equation}
while the base duration across the retained serial module boundaries is
\begin{equation}
C_{\rm LS}=\sum_a n_a C_{\rm LS}(M_a).
\end{equation}
Within a family, explicit quantum and classical dependencies determine local overlap under the fixed-area constraint. A single inventory replay of the composed T requests supplies $C_{\rm wait}$, yielding $C_{\rm exec}=C_{\rm LS}+C_{\rm wait}$. Inventory and factory phase are retained across repeated calls, so each module's attributed execution time depends on the shared global configuration. Data, buffer, and factory footprints are provisioned once for the complete execution.

The hierarchical evaluation combines local LS profiles with the declared upper-level ordering and area assumptions. 

Additional execution-model derivations, benchmark definitions, architecture-search details, resource-model parameters, cultivation models, and extended results are provided in the Supplemental Material~\cite{SupplementalMaterial}.

\section*{Data Availability}

The source data underlying the figures and tables, including the
benchmark records and layout-comparison results, are available from the
corresponding authors upon reasonable request.

\section*{Code Availability}

The compiler, architecture optimizer, fault-tolerant schedule generator,
and scripts used to reproduce the reported estimates are available from
the corresponding authors upon reasonable request.

\section*{Author Contributions}

X.Y., Ying Li, and Y.Y. conceived the project.
J.H. and Y.Y. developed the hierarchical decomposition framework and designed the algorithms.
Y.T., Z.H., and Y.Y. developed the software for workload optimization, compilation, and resource estimation.
Z.Y., Z.W., Z.D., Y.Z., Yangyu Lu, Z.Z., H.X., and Y.H. contributed to the analysis and evaluation of logical operations, surface codes, layout design, and decoding.
Z.H., Y.T., and J.H. analyzed and interpreted the final results.
Y.Y., X.Y., and J.S. supervised the project.
All authors contributed to writing, reviewing, and revising the manuscript.

\section*{Conflict of Interest}

The authors declare no competing interests.


\clearpage

\setcounter{section}{0}
\setcounter{subsection}{0}
\setcounter{subsubsection}{0}
\setcounter{equation}{0}
\setcounter{figure}{0}
\setcounter{table}{0}

\renewcommand{\thesection}{S\arabic{section}}
\renewcommand{\thesubsection}{\thesection.\arabic{subsection}}
\renewcommand{\thesubsubsection}{\thesubsection.\arabic{subsubsection}}
\renewcommand{\theequation}{S\arabic{equation}}
\renewcommand{\thefigure}{S\arabic{figure}}
\renewcommand{\thetable}{S\arabic{table}}
\renewcommand{\theHequation}{supp.equation.\arabic{equation}}
\renewcommand{\theHfigure}{supp.figure.\arabic{figure}}
\renewcommand{\theHtable}{supp.table.\arabic{table}}
\makeatletter
\renewcommand{\p@subsection}{}
\renewcommand{\p@subsubsection}{}
\makeatother

\section*{Supplementary Information}

This Supplementary Information provides implementation details that are omitted from the main text for brevity. We focus on the construction of the twenty benchmark circuits, the relation between the optimization stages used in the algorithm-to-execution workflow, the circuit-specific layout search, the non-Clifford resource model and optimization procedure, and the numerical reconstruction of the application-scale secp256k1 example. Additional technical tables collect auxiliary assumptions, including decoder timing and resource-state factory parameters.

\section{Benchmark circuits and executable instances}
\label{sec:sm_benchmarks}

\subsection{Benchmark set}

The benchmark suite contains twenty explicit logical circuits spanning seven algorithm families: amplitude estimation, Grover search, quantum approximate optimization, quantum linear-system algorithms, quantum phase estimation, digital quantum simulation, and Shor-type algorithms. The purpose of this set is not to provide an exhaustive taxonomy of quantum algorithms, but to sample circuits with different interaction patterns, logical parallelism, and non-Clifford demand.

Each benchmark is instantiated as a concrete logical circuit before fault-tolerant analysis. The main suite uses the twenty supplied ZX/frame-transformed inputs dated 2026-09-28; the earlier pre-ZX circuits are not used as the current numerical baseline. The input operation set is the axis-aware set in Eq.~\eqref{eq:rz-gate-set}, including $R_X$ and $R_Z$ with $R_P(\varphi)=e^{-i\varphi P/2}$. Resource compilation retains axes and available measurement metadata, expands selected rotations, and reconstructs dependencies. The resource adapter does not claim to implement arbitrary dynamic classical control.
\begin{equation}
\mathcal G_{\rm input}=\mathcal G\text{ of Eq.~\eqref{eq:rz-gate-set}}.
\label{eq:sm_gate_set}
\end{equation}
The additional twelve QAOA readout-aware inputs comprise checkpoints 000, 020, 040, 060, 080, and 100 for AA8 and KSAT12, dated 2026-09-29. Their measurement conventions differ from the main-suite instances, so train100 does not replace either main benchmark. These samples are evaluated per circuit execution; training iteration counts and measurement-shot weights are not inferred.

\subsection{Logical scheduling}

The logical schedule is constructed from the operation-dependency graph. Let $Q(G_i)$ denote the logical-qubit support of operation $G_i$, and let $G_i\prec G_j$ denote an explicit precedence relation. Two operations may occupy the same logical layer only when
\begin{equation}
Q(G_i)\cap Q(G_j)=\varnothing,
\qquad
G_i\nprec G_j,
\qquad
G_j\nprec G_i.
\label{eq:sm_schedule_condition}
\end{equation}
For the explicit benchmark resource adapter, the graph preserves the supplied gate order on every shared logical qubit. The broader hierarchical representation can carry dynamic-event dependencies, but the explicit resource adapter has not been validated as a general conditional-circuit scheduler.

Resource-specific synthesis and frame propagation first produce a new gate stream. Earliest dependency layers are rebuilt without changing shared-qubit order. Physical scheduling subsequently resolves routing, patch, and Rz-slot conflicts; the legalizer splits rejected concurrent layers while preserving gate order. Logical depth and duration-weighted LS depth are consequently different quantities.

The three descriptors passed to the subsequent optimization are
\begin{equation}
\mathcal{W}
=
\left\{
G_{\mathrm{int}},
P(\ell),
r_{\mathrm{NC}}(\ell)
\right\},
\label{eq:sm_workload}
\end{equation}
where $G_{\mathrm{int}}$ is the weighted logical interaction graph, $P(\ell)$ records the number of operations assigned to logical layer $\ell$, and $r_{\mathrm{NC}}(\ell)$ records the non-Clifford demand in that layer.

\subsection{Benchmark statistics and validation}

For every explicit benchmark, the normalized circuit and scheduled representation are checked against the original generated instance. The validation includes logical width, gate counts, measurement and reset counts, logical-layer count, number of rotations, and layer-resolved non-Clifford events.

\begin{table*}[htbp]
\centering
\scriptsize
\caption{Logical statistics of the current twenty ZX/frame inputs before resource-specific synthesis. Layers are reconstructed from shared-qubit dependencies; peak rotation counts include both axes. Residual H counts are input operations, not automatically charged physical H gates.}
\label{tab:benchmark_statistics}
\resizebox{\textwidth}{!}{%
\begin{tabular}{llrrrrrrr}
\toprule
Family & Circuit & Qubits & Gates & Layers & $R_X$ & $R_Z$ & H & Peak rotations \\
\midrule
Amplitude estimation & AE--BHMT & 8 & 2,829 & 2,419 & 122 & 613 & 0 & 6 \\
Amplitude estimation & AE--pricing & 8 & 4,889 & 4,532 & 433 & 1,091 & 0 & 3 \\
Amplitude estimation & AE--power & 12 & 12,630 & 11,775 & 98 & 1,729 & 0 & 9 \\
Grover search & Grover--SAT & 11 & 454 & 267 & 96 & 114 & 4 & 2 \\
Grover search & Grover--AES & 15 & 502 & 269 & 96 & 114 & 4 & 2 \\
Grover search & Grover--hash & 15 & 508 & 267 & 96 & 114 & 4 & 2 \\
QAOA & QAOA--AA8 & 8 & 1,928 & 1,731 & 33 & 279 & 8 & 5 \\
QAOA & QAOA--KSAT12 & 12 & 3,610 & 3,318 & 510 & 12 & 12 & 12 \\
QAOA & QAOA--MaxCut & 12 & 144 & 81 & 36 & 24 & 12 & 4 \\
QLSA & QLSA--HHL & 10 & 928 & 625 & 376 & 96 & 5 & 3 \\
QLSA & QLSA--QSVT & 8 & 231 & 104 & 52 & 132 & 0 & 3 \\
QPE & QPE--factor & 15 & 17,181 & 13,607 & 3,415 & 3,503 & 0 & 5 \\
QPE & QPE--state prep. & 15 & 4,553 & 3,386 & 619 & 1,103 & 0 & 6 \\
QPE & QPE--Trotter & 15 & 4,133 & 3,268 & 559 & 983 & 0 & 9 \\
Quantum simulation & QSim--JW & 12 & 322 & 204 & 28 & 74 & 0 & 4 \\
Quantum simulation & QSim--tiled & 12 & 322 & 207 & 28 & 74 & 0 & 4 \\
Quantum simulation & QSim--Heisenberg & 12 & 208 & 66 & 32 & 80 & 0 & 6 \\
Shor & Shor--binary ECDLP & 12 & 4,293 & 2,782 & 1,324 & 838 & 0 & 2 \\
Shor & Shor--prime ECDLP & 16 & 5,851 & 3,797 & 1,752 & 1,180 & 0 & 2 \\
Shor & Shor--RSA15 & 10 & 4,465 & 2,939 & 1,420 & 838 & 0 & 3 \\
\bottomrule
\end{tabular}%
}
\end{table*}

\section{Optimization workflow and relation between objectives}
\label{sec:sm_workflow}

Logical compilation, architecture synthesis, and resource estimation remain separate modules connected by explicit interfaces. The resource experiment makes repeated architecture/schedule calls for candidate rotation implementations and then evaluates their physical costs. The distinction is between an inner layout search and an outer physical-resource ranking, not a single irreversible selection of layout before all resource decisions.

\begin{figure*}[htbp]
\centering
\includegraphics[width=\textwidth]{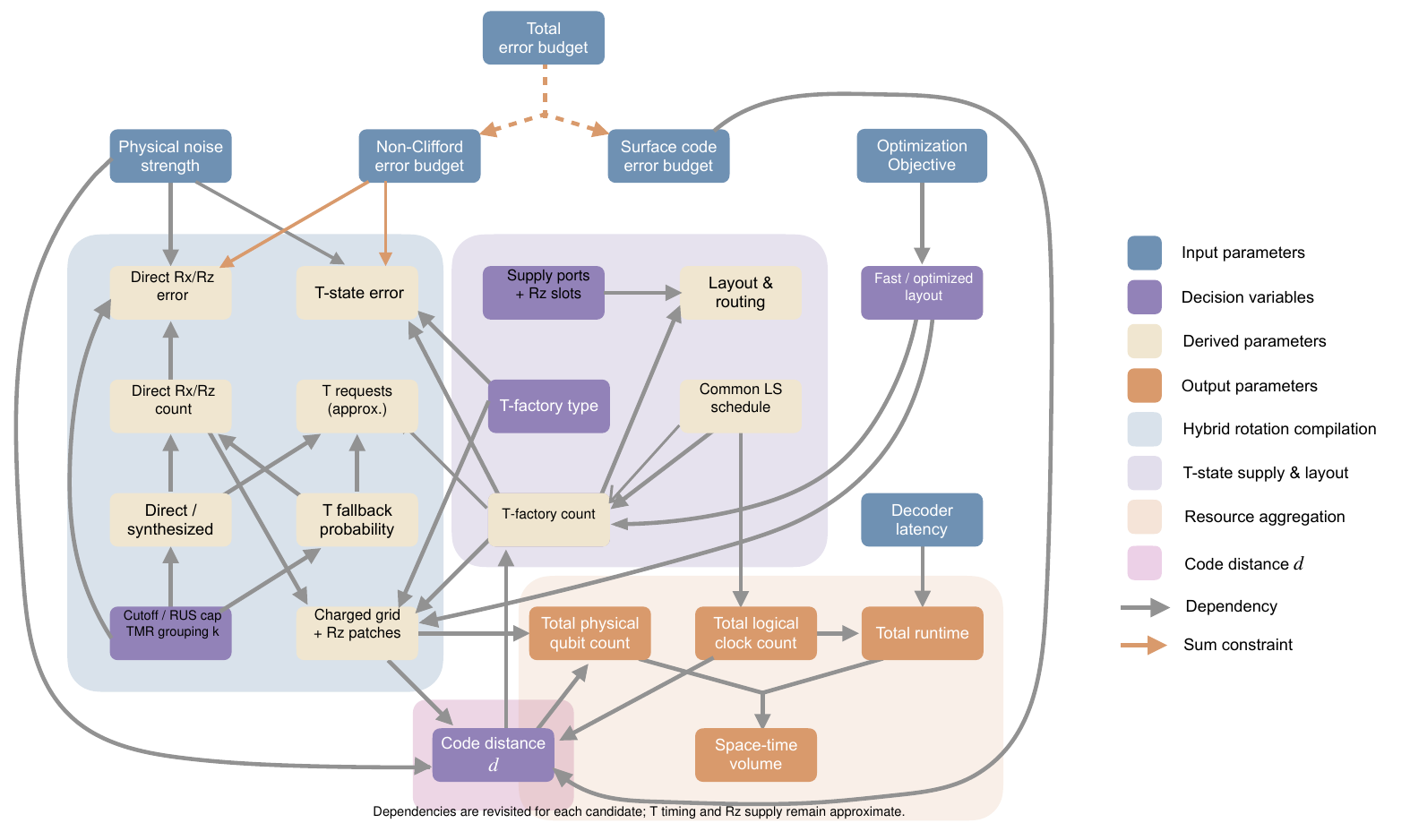}
\caption{\textbf{Candidate evaluation across the algorithm-to-execution workflow.} The diagram retains the input, decision, derived-quantity, and resource-output regions. Arrows denote dependencies revisited during candidate evaluation, not a one-pass execution order. The common LS scheduler evaluates fast or circuit-specific layouts with Rz slot limits and full conditional fallback mean time. Approximate T request events and probability-weighted fallback demand determine factory multiplicity through the demand window. Both error budgets and physical cost are recomputed while comparing rotation policies, TMR grouping $k$, odd $d$, factory types, Rz capacity, and layouts. Perimeter access locations represent supply interfaces; factory bodies are charged separately. The diagram does not imply calibrated accepted-state supply or certified global optimality.}
\label{fig:sm_workflow}
\end{figure*}

For a compiled structure $s$, the architecture module supplies a finite candidate set
\begin{equation}
\mathcal C_{\mathcal A}(s)=\mathrm{LayoutSearch}(s;\text{grid, slots, seed, budget}).
\label{eq:sm_layout_selection}
\end{equation}
The physical evaluator selects
\begin{equation}
\widehat x=\arg\min_{x\in\mathcal C_{\rm evaluated},\,\mathrm{feasible}(x)}V_{\rm ST}(x),
\label{eq:sm_physical_selection}
\end{equation}
where the resource dimensions include threshold or RUS cap, $k$, odd $d$, factory type, multiplicity derived from demand, Rz capacity, and the returned layout. Factory multiplicity is not an unrelated free parameter: it is recomputed by Eq.~\eqref{eq:factory-number} for each candidate.
\begin{equation}
V_{\rm ST}=N_{\rm phys}\lceil\bar C_{\rm sched}\rceil d.
\label{eq:sm_spacetime}
\end{equation}
The reported corrected benchmark search includes odd $d=5,7,\ldots,37$ and compatible $k=2,\ldots,\lfloor d/2\rfloor$. TMR uses data distance as its grouping bound, while factory distances are catalogue parameters. Candidates violating either $10^{-2}$ budget are rejected; a layout-independent lower bound prunes policies already infeasible even with the best accepted-state factory error. Equivalent compiled structures reuse work. Several structures and geometries receive refinement, but the finite candidate set still yields best-found feasible values rather than global minima.

\section{Genetic optimization of circuit-specific layouts}
\label{sec:sm_genetic}

\subsection{Layout variables}

A candidate layout specifies the assignment of logical qubits to data-patch locations and the orientation of each logical patch. Patch orientation determines which $X$ or $Z$ boundary is exposed along each grid direction,
\begin{equation}
(\ell_{\mathrm{lr}},\ell_{\mathrm{ud}})
\in
\{X,Z\}^2,
\qquad
\ell_{\mathrm{lr}}\neq\ell_{\mathrm{ud}}.
\end{equation}

For logical qubits $i$ and $j$, the communication term is
\begin{equation}
C_{\mathrm{CNOT}}
=
\sum_{i\neq j}
C_{ij}d(i,j),
\label{eq:sm_cnot_cost}
\end{equation}
where $C_{ij}$ is the interaction count and $d(i,j)$ is the shortest compatible routing distance.

The non-Clifford access term is
\begin{equation}
C_T
=
\sum_i T_i d(i),
\label{eq:sm_resource_access}
\end{equation}
where $d(i)$ is the distance from data patch $i$ to the nearest resource-state access perimeter.

Boundary compatibility is represented by $C_H$ and $C_S$, the supplied backend's access penalties for the required Hadamard and phase operations. These terms depend on the assigned patch orientations and surrounding routing geometry; the resource wrapper retains the backend's evaluation rather than introducing a separate closed-form penalty function.

The static objective is
\begin{equation}
C_{\mathrm{layout}}
=
C_{\mathrm{CNOT}}+C_T+C_H+C_S,
\label{eq:sm_layout_objective}
\end{equation}
which expresses the architecture-side proxy with unit weights. Resource evaluation does not reinterpret this proxy as physical volume or change the supplied optimizer's internal search operators. The numerical wrapper settings used for this revision are specified below.

\subsection{Search procedure}

The resource wrapper uses the supplied \texttt{qec\_est} backend without modifying its internal genetic operators. For logical width $n$ and fast-template area $A_f$, it screens every integer rectangular interior $(a,b)$ satisfying $n\leq ab\leq2A_f$, including both orientations and narrow rectangles, plus the previous search dimensions. For each integer Rz slot count from one to the maximum rebuilt-layer direct-rotation concurrency, a short operation-coverage probe is optimized at seed 11, five iterations, and initial/population sizes 8/8. The four lowest probe-volume proxies, the fastest candidate, and the smallest-area candidate form the shape shortlist, with duplicates removed. Every retained compiled policy is subsequently evaluated on applicable shortlisted layouts. This screening is not exhaustive full-circuit optimization over every rectangle.

For each of All-$T$, fixed threshold, and fixed RUS cap, up to two best compiled structures and two geometries per structure receive full-input refinement at seeds 11, 29, and 47, 50 iterations, and initial/population sizes 20/20. The fast template is evaluated in both orientations, initially at seed 29 with seeds 11/47 available on failure. Saved feasible candidates remain eligible. No selected optimized plan reaches the searched upper area bound, although this does not prove that larger or differently initialized layouts could not improve it.

All feasible fast plans are admitted to the corresponding optimized pool, and the All-$T$ policy is admitted to Hybrid. Thus selected optimized volume cannot exceed selected fast volume, and Hybrid volume cannot exceed its All-$T$ control. Five main-suite workloads have identical zero-direct-rotation compiled inputs in both policies and share feasible candidates. Policy structures with identical compiled inputs and error/demand data are deduplicated.

The twelve QAOA checkpoints use the same four-arm procedure, rectangle screening, and refinement settings as the twenty main benchmarks. Each remains an independent input with its own selected parameters. Source hashes, candidate records, selected layout recipes, lowered circuits, and fresh-object replay checks are retained for all 32 inputs.

\subsection{Routed evaluation}

The backend returns LS layers as a T count and duration. Cumulative durations define layer starts, and each count contributes that many approximate request events. These are not actual injection timestamps: even serial T operations can be reported together at an earlier layer. The legalizer queries the LS engine to split conflicts into ordered feasible subsets, preserving dependencies. Long inputs can be repaired in bounded windows and then rechecked as a whole; controls are not unconditionally serialized merely because a payload is large. The gate-level fallback API is not used as evidence of physical routability.

The fast template~\cite{Litinski2019gameofsurfacecodes} and optimized layouts use the same LS engine, perimeter access convention, and local repair rules. If a rejected control $C(P_i,Q_j)$ has no accepted direct realization, an existing data qubit $m$ may mediate it. With $A=C(P_i,X_m)$ and $B=C(Z_m,Q_j)$, the chronological sequence $A,B,A,B$ implements $C(P_i,Q_j)\otimes I_m$ for arbitrary mediator input. Both mediator axes and available data qubits are tested; the lowest accepted isolated LS duration is used. The full matrix identity is checked for the endpoint Pauli combinations, and the complete lowered circuit is replayed on a fresh layout. Extra control time and mediator dependencies are charged, with no extra T state or patch. This conservative alternative does not establish a shortest route or prove that the rejected direct operation is physically impossible. A candidate with no verified alternative is rejected.

The common adapter propagates frames before handoff, omits zero-time Pauli updates and terminal measurements from LS timing, and appends a three-clock terminal layer when quantum-output H corrections remain. Input measurements are terminal on their qubits, so readout can be deferred; the exposure model retains the data patches throughout. Both layout arms therefore share endpoint conventions. Parameters can still change between plans, so this is not a frozen-resource routing ablation.

\section{Non-Clifford resource model and optimization}
\label{sec:sm_nonclifford}

\subsection{Rotation preprocessing}

For an input angle $\varphi$, set $m=\operatorname{round}(2\varphi/\pi)$ and $\beta=\varphi-m\pi/2$. The signed residual and Clifford factor are both retained:
\begin{equation}
R_P(\varphi)=e^{-im\pi/4}S_P^mR_P(\beta),\qquad |\beta|\leq\pi/4.
\label{eq:sm_angle_reduction}
\end{equation}
Here $S_P=e^{i\pi/4}R_P(\pi/2)$ is the phase Clifford about axis $P$. Exact multiples of $\pi/4$ use exact Clifford+$T$ words and zero synthesis error. The input and exported angles use the standard half-angle convention; STAR's $e^{i\theta_L P}$ convention uses signed $\theta_L=-\beta/2$, with magnitude used for error evaluation. Threshold values in resource records use standard radians; a paper-convention threshold is half that value.

Threshold candidates include zero, endpoint values, every $2^j|\beta|\leq\pi/2$, and values immediately above each switching boundary. A supplementary $\pi/96$ mesh is retained, but is not the resolution limit. Equivalent trial-count profiles are deduplicated and clearly infeasible analog-error candidates are pruned for each $k$; the union over allowed $k$ is searched. Thus even very small input-dependent cutoffs are represented.

\subsection{\texorpdfstring{Clifford+$T$}{Clifford+T} branch}

The implementation uses \texttt{pygridsynth} 2.0.0, seed 0, and stores the actual execution-ordered gate word, scalar phase, tolerance, and independently verified operator-norm error~\cite{RossSelinger2014optimal}. If $M_{\rm app}$ is the expected number of non-exact approximation events, allocate
\begin{equation}
\epsilon_{\rm word}=2^{\lfloor\log_2(10^{-4}/\max(1,M_{\rm app}))\rfloor}.
\label{eq:sm_synthesis}
\end{equation}
Deterministic words have weight one and fallback words their branch probability in the synthesis-error allocation. This is the declared additive synthesis-error proxy within the non-Clifford budget, not a sampled failure rate.

Frames are propagated through the generated words. Pauli gates take zero time; surviving $S/S^\dagger$ operations take one clock, T operations require one injection clock plus any separately scheduled boundary preparation, and unavoidable physical H takes three. The resource handoff keeps retained rotations with conditional fallback metadata rather than sampling or expanding random branches into the deterministic logical circuit.

Expected T consumption is
\begin{align}
\bar N_T={}&N_T^{\rm original}+N_T^{\rm exact}
+N_T^{\rm synthesized}\nonumber\\
&+\sum_i p_iN_{T,i}^{\rm fallback}.
\end{align}
The exact term counts T gates needed by rotations exactly on the digital grid, whereas the synthesized term counts approximate non-grid rotations. These categories are disjoint. Every contribution uses the same selected factory protocol.

\subsection{Direct rotation branch}

The direct branch uses STAR-magic mutation~\cite{Toshio2026starmagicmutation}. For target magnitude $\theta_L=|\beta|/2$ and grouping count $k$, numerically invert the small-angle TMR branch to obtain the physical rotation $u$:
\begin{align}
p_{\rm ideal}&=\sin^{2k}u+\cos^{2k}u,\\
\theta_L&=\arcsin\!\left(\frac{\sin^ku}{\sqrt{p_{\rm ideal}}}\right).
\end{align}
Define the leading erroneous branch by
\begin{align}
p_{\rm err}&=\sin^2u\cos^2u\left(\sin^{2k-4}u+\cos^{2k-4}u\right),\\
\theta_{\rm err}&=-\arcsin\!\left(\frac{\sin^{k-1}u\cos u}{\sqrt{p_{\rm err}}}\right),\\
\bar q_1&\simeq\frac{p_{\rm err}}{p_{\rm ideal}}\frac{k}{15}p_{\rm phys},\\
e(\theta_L,k)&=2\bar q_1\sin^2(\theta_{\rm err}-\theta_L).
\end{align}
The main configuration assumes the leading native-realization coefficient $P_{\rm ud}=(k/15)p_{\rm phys}$. The denominator 15 is the number of nonidentity two-qubit Pauli errors in a uniform depolarizing model. A decomposed hardware realization can have a different coefficient; alternative coefficient and partition conventions are sensitivity assumptions rather than hardware calibration. For partitions into groups of at least two qubits, the model searches $2\leq k\leq\lfloor d/2\rfloor$; it does not simulate each hardware decomposition or its higher-order noise.

The implemented leading-order analog contribution includes the accumulated trial errors on both successful and all-failed RUS paths:
\begin{equation}
\begin{aligned}
P_{{\rm analog},i}&=\sum_{m=1}^{L_i}2^{-m}\sum_{j=0}^{m-1}e(2^j\theta_{L,i},k)\\
&\quad+2^{-L_i}\sum_{j=0}^{L_i-1}e(2^j\theta_{L,i},k)\\
&=\sum_{j=0}^{L_i-1}2^{-j}e(2^j\theta_{L,i},k).
\end{aligned}
\label{eq:sm_direct_error}
\end{equation}
and $\alpha_{{\rm RUS},i}=P_{{\rm analog},i}/(\theta_{L,i}p_{\rm phys})$. The all-failed term retains the analog noise accumulated before digital fallback. Digital synthesis error and factory-state error are charged separately. This angle- and policy-dependent expression is a leading-order conditional estimate with $p_{\rm suc}$ approximated by $p_{\rm ideal}$; it is not a calibrated finite-noise preparation simulation.

The threshold policy doubles the signed residual until it reaches the cutoff or an exact digital angle. The fixed-RUS policy allows at most $N=1,\ldots,8$ trials, stopping earlier if the next TMR target is outside $|\theta_L|\leq\pi/4$. Thus the effective $L_i$ can depend on the angle even at fixed $N$. Its fallback correction has standard angle $2^{L_i}\beta_i$ and probability $p_i=2^{-L_i}$. Both policies are evaluated rather than replacing one with the other.

The analog duration is $\bar n_i+1$ clocks, with $\bar n_i=2(1-2^{-L_i})$ and one fixed execution allowance. For six trials this is $2.96875$ clocks under pre-supply. The complete conditional Clifford+$T$ fallback is evaluated using the reference fast primitive timing, including non-Pauli Clifford and physical H costs; $p_i C_{{\rm fb},i}$ is added to the parent rotation duration in both layout arms. A reference fallback T at offset $\delta$ has event time $t_i+\bar n_i+1+\delta$ and weight $p_i$: probability scales its weight, not its offset. Reference fallback-event times are then rescaled to the returned LS duration. This includes the full digital word but does not explicitly route stochastic fallback gates on each selected geometry or reproduce a stochastic makespan.

Both fast and optimized layouts use the same integer Rz slot limit. R0 charges one patch per slot and R1 three; zero direct rotations require no charged Rz fleet even if the backend contains an unused slot marker. R0 retains its historical three-clock supply label, but no separate three- or four-clock Rz-demand sliding window is applied in this common-LS run. Slots constrain concurrent complete rotations carrying their mean execution costs; they do not certify a three-clock accepted-state production rate. Ideal $p_{\rm ideal}$ is diagnostic and does not alter supply. R1 reselects feasible resource parameters over the same layout pool. An angle-dependent accepted-state rate remains unevaluated.

\subsection{Magic-state error and rejection}

Postselected preparation of arbitrary-rotation resource states motivates treating accepted-state infidelity separately from heralded preparation failure~\cite{choi2023fault}.
For positive expected demand, the target accepted-state error is
\begin{equation}
p_T^{\rm tar}=\frac{\epsilon_{\rm NC}-\sum_iP_{{\rm analog},i}-P_{\rm syn}}{\bar N_T}.
\label{eq:sm_magic_target}
\end{equation}
A candidate is infeasible if this remaining budget is negative. The magic-state output contribution $\bar N_Tp_T^{\rm out}$ and the surface-code exposure of provisioned factory patches are distinct terms in the stated accounting convention.

Two different failure probabilities are treated separately. The residual error $p_T^{\mathrm{out}}$ of an accepted state is unheralded and contributes to the non-Clifford error budget. The discard probability $q$ is heralded and affects preparation latency and throughput instead.

For a single postselected candidate,
\begin{equation}
p_{\mathrm{succ}}=1-q,
\end{equation}
and
\begin{equation}
\mathbb{E}[N_{\mathrm{attempt}}]
=
\frac{1}{1-q}.
\end{equation}

\subsection{Factory throughput sizing}

Let $(t_a,w_a)$ be the approximate deterministic T-request histogram and probability-weighted fallback-event trace described above. For the selected factory's expected output period $c_F$, use the half-open continuous window
\begin{equation}
N_F=\left\lceil\max_t\sum_{a:\,t\leq t_a<t+c_F}w_a\right\rceil.
\label{eq:sm_factory_number}
\end{equation}
There is no separate integer rounding of the window width or substitution of circuit-average fallback demand. Numerical implementation uses a $10^{-10}$ tolerance before the ceiling. Zero demand permits zero factories. Each candidate uses one factory type for all T origins, with its multiplicity recalculated after scheduling. No preparation startup cost or explicit stochastic inventory/waiting simulation is included.

\subsection{Pre-patch parallel cultivation}
\label{subsec:sm_ppc}

Pre-patch parallel cultivation (PPC) is illustrated conceptually by Fig.~\ref{fig:magic}. Its numerical resource model uses the finite protocol geometry
\begin{equation}
N_S=\left\lfloor\frac{D_2}{D_1}\right\rfloor^2,
\end{equation}
where $D_1$ is the small cultivated code distance and $D_2$ its final grafting distance, not the data-code distance. Independent candidates with overall discard rate $q$ give
\begin{equation}
P_{\rm survive}=1-q^{N_S}.
\label{eq:sm_ppc_survival}
\end{equation}
For batch-attempt duration $T_{\rm attempt}$,
\begin{equation}
T_{\rm eff}=\frac{T_{\rm attempt}}{1-q^{N_S}}.
\label{eq:sm_ppc_latency}
\end{equation}
Because the catalogue periods $c_{F,\rm cat}$ already include single-candidate repetition, the period used in the demand window is
\begin{equation}
c_{F,\rm PPC}=c_{F,\rm cat}\frac{1-q}{1-q^{N_S}}.
\end{equation}
This avoids applying the original discard penalty twice. The matched geometries $(D_1,D_2,N_S)$ are Gidney $f=3:(3,15,25)$ and $f=5:(5,15,9)$; Folded-H $f=3:(3,13,16)$ and $f=5:(5,15,9)$; RP$^2$ and Claes $f=3:(3,7,4)$ and $f=5:(5,11,4)$. The $10^{-4}$ ungrown entry has no matched end-to-end PPC adjustment.

The approximation applies the full-protocol discard rate to independent simultaneous candidates and retains the catalogue factory footprint. It does not resolve intermediate growth failures, simultaneous survivor growth, collision-free grafting, or a spatially simulated processor-wide seed farm. The ideal geometric figures motivate the proposal but do not validate these missing stages. Stored no-PPC factory counts are diagnostics on a fixed demand trace, not an independently reoptimized no-PPC experiment.

\subsection{Circuit-resolved hybrid and layout results}

The tables below use the complete 2026-10-09 reachable-RUS-history correction and subsequent factory-catalogue control for all twenty main circuits and twelve independent QAOA checkpoints. The corrected stage reselects distance, grouping, and factories within the retained verified candidate pool; it does not add an exhaustive geometry search. All-$T$ fast is the common baseline. The main gain matrix separates Hybrid alone, layout alone under All-$T$, and their combination, while also retaining the Hybrid-conditioned layout and optimized-pool Hybrid controls. T-event placement, mean execution time, and accepted-state supply retain the approximations stated above; no in-progress stochastic-batch results are substituted.

\begin{table*}[htbp]
\centering
\scriptsize
\caption{Corrected gain matrix. $T_f,H_f,T_o,H_o$ denote All-$T$/Hybrid on fast/optimized-pool layouts. The first three columns use the common All-$T$ fast baseline; the last two give conditional layout and Hybrid controls. All ratios are volume gains, independently reselected within the retained verified candidate pool under the same stated model.}
\label{tab:hybrid-synthesis-comparison}
\resizebox{0.6\textwidth}{!}{%
\begin{tabular}{lrrrrr}
\toprule
Circuit & $T_f/H_f$ & $T_f/T_o$ & $T_f/H_o$ & $H_f/H_o$ & $T_o/H_o$ \\
\midrule
AE--BHMT & 2.7446 & 1.0000 & 2.8342 & 1.0327 & 2.8342 \\
AE--pricing & 2.0677 & 1.0000 & 2.1599 & 1.0446 & 2.1599 \\
AE--power & 7.2395 & 1.0000 & 8.6077 & 1.1890 & 8.6077 \\
Grover--SAT & 1.0000 & 1.5461 & 1.5461 & 1.5461 & 1.0000 \\
Grover--AES & 1.0000 & 1.2779 & 1.2779 & 1.2779 & 1.0000 \\
Grover--hash & 1.0000 & 1.2858 & 1.2858 & 1.2858 & 1.0000 \\
QAOA--AA8 & 11.2689 & 1.0000 & 12.7048 & 1.1274 & 12.7048 \\
QAOA--KSAT12 & 10.3504 & 1.0000 & 11.3263 & 1.0943 & 11.3263 \\
QAOA--MaxCut & 1.8096 & 1.3001 & 1.9462 & 1.0755 & 1.4969 \\
QLSA--HHL & 9.2493 & 1.1604 & 11.1373 & 1.2041 & 9.5979 \\
QLSA--QSVT & 5.6231 & 1.4181 & 6.9143 & 1.2296 & 4.8758 \\
QPE--factor & 1.0910 & 1.0000 & 1.0910 & 1.0000 & 1.0910 \\
QPE--state prep. & 1.6556 & 1.0007 & 1.6556 & 1.0000 & 1.6545 \\
QPE--Trotter & 2.0191 & 1.0000 & 2.1835 & 1.0814 & 2.1835 \\
QSim--JW & 5.7878 & 1.0000 & 6.6699 & 1.1524 & 6.6699 \\
QSim--tiled & 5.7856 & 1.0000 & 6.3210 & 1.0925 & 6.3210 \\
QSim--Heisenberg & 5.0493 & 1.2655 & 6.4402 & 1.2755 & 5.0889 \\
Shor--binary ECDLP & 1.0000 & 1.1155 & 1.1155 & 1.1155 & 1.0000 \\
Shor--prime ECDLP & 1.0000 & 1.0426 & 1.0426 & 1.0426 & 1.0000 \\
Shor--RSA15 & 1.1666 & 1.1479 & 1.3232 & 1.1342 & 1.1528 \\
\bottomrule
\end{tabular}%
}
\end{table*}
\begin{table*}[htbp]
\centering
\scriptsize
\caption{Selected Hybrid fast-block plans from the corrected records. Thresholds are standard rotation radians (twice the STAR convention); $N$ is the nominal RUS cap. $k$ is the grouping count and $N_R$ the charged direct-rotation patches. Expected T demand includes fallback. FH3/FH5, RP3/RP5, CL3/CL5, and GI3/GI5 identify Folded-H, RP$^2$, Claes, and Gidney protocols with fault distance 3/5, followed by multiplicity. Times are seconds and volumes physical-qubit--stabilizer-rounds; optimized-pool plans can retain fast.}
\label{tab:hybrid-fast-plans}
\resizebox{0.88\textwidth}{!}{%
\begin{tabular}{llrrrrrrrr}
\toprule
Circuit & Policy & $k$ & $d$ & Factory:count & $N_R$ & $\bar N_T$ & $Q$ & $t$ & $V$ \\
\midrule
AE--BHMT & $\theta=0.28982$ & 2 & 17 & FH5:27 & 1 & 9659.50 & 21,509 & 0.14722 & $6.3331\times10^{9}$ \\
AE--pricing & $\theta=0.126303$ & 2 & 17 & FH5:25 & 1 & 19826.50 & 21,027 & 0.3160555 & $1.3291\times10^{10}$ \\
AE--power & $\theta=0.321751$ & 3 & 17 & FH5:29 & 1 & 9690.62 & 27,761 & 0.2792165 & $1.5503\times10^{10}$ \\
Grover--SAT & $\theta=0$ & 2 & 13 & RP3:7 & 0 & 210.00 & 12,516 & 0.002704 & $6.7687\times10^{7}$ \\
Grover--AES & $\theta=0$ & 2 & 13 & RP3:7 & 0 & 210.00 & 15,886 & 0.0026585 & $8.4466\times10^{7}$ \\
Grover--hash & $\theta=0$ & 2 & 13 & RP3:7 & 0 & 210.00 & 15,886 & 0.002574 & $8.1781\times10^{7}$ \\
QAOA--AA8 & $\theta=0.392699$ & 3 & 15 & RP3:17 & 1 & 2751.47 & 13,425 & 0.0362475 & $9.7325\times10^{8}$ \\
QAOA--KSAT12 & $\theta=0.322448$ & 3 & 15 & FH5:11 & 1 & 2424.00 & 18,815 & 0.070275 & $2.6444\times10^{9}$ \\
QAOA--MaxCut & $\theta=0.940567$ & 2 & 15 & FH5:13 & 3 & 1362.00 & 20,195 & 0.0135225 & $5.4617\times10^{8}$ \\
QLSA--HHL & $\theta=1.1781$ & 2 & 13 & RP3:7 & 1 & 436.75 & 11,505 & 0.006344 & $1.4598\times10^{8}$ \\
QLSA--QSVT & $\theta=1.0472$ & 2 & 13 & RP3:7 & 1 & 390.00 & 9,483 & 0.005252 & $9.9609\times10^{7}$ \\
QPE--factor & $\theta=0.173036$ & 2 & 19 & FH5:15 & 1 & 222814.50 & 36,781 & 3.2406495 & $2.3839\times10^{11}$ \\
QPE--state prep. & $\theta=0.141897$ & 2 & 17 & FH5:41 & 1 & 42090.00 & 36,423 & 0.50456 & $3.6755\times10^{10}$ \\
QPE--Trotter & $N=1$ & 3 & 17 & FH5:9 & 1 & 25053.50 & 28,711 & 0.3762355 & $2.1604\times10^{10}$ \\
QSim--JW & $\theta=0.8$ & 2 & 13 & RP3:6 & 3 & 224.00 & 13,424 & 0.0043745 & $1.1745\times10^{8}$ \\
QSim--tiled & $\theta=0.8$ & 2 & 13 & RP3:6 & 3 & 224.00 & 13,424 & 0.0043745 & $1.1745\times10^{8}$ \\
QSim--Heisenberg & $\theta=0.8$ & 2 & 13 & RP3:5 & 3 & 336.00 & 13,321 & 0.0042315 & $1.1274\times10^{8}$ \\
Shor--binary ECDLP & $\theta=0$ & 2 & 15 & RP3:6 & 0 & 2162.00 & 16,333 & 0.0270825 & $8.8468\times10^{8}$ \\
Shor--prime ECDLP & $\theta=0$ & 2 & 15 & RP3:6 & 0 & 2932.00 & 20,823 & 0.036225 & $1.5086\times10^{9}$ \\
Shor--RSA15 & $\theta=0.392699$ & 2 & 15 & RP3:6 & 1 & 2252.75 & 14,986 & 0.028605 & $8.5735\times10^{8}$ \\
\bottomrule
\end{tabular}%
}
\end{table*}
\begin{table*}[htbp]
\centering
\scriptsize
\caption{Selected Hybrid optimized-pool plans from the corrected records. Thresholds are standard rotation radians (twice the STAR convention); $N$ is the nominal RUS cap. $k$ is the grouping count and $N_R$ the charged direct-rotation patches. Expected T demand includes fallback. FH3/FH5, RP3/RP5, CL3/CL5, and GI3/GI5 identify Folded-H, RP$^2$, Claes, and Gidney protocols with fault distance 3/5, followed by multiplicity. Times are seconds and volumes physical-qubit--stabilizer-rounds; optimized-pool plans can retain fast.}
\label{tab:hybrid-optimized-plans}
\resizebox{0.89\textwidth}{!}{%
\begin{tabular}{llrrrrrrrr}
\toprule
Circuit & Policy & $k$ & $d$ & Factory:count & $N_R$ & $\bar N_T$ & $Q$ & $t$ & $V$ \\
\midrule
AE--BHMT & $\theta=0.28982$ & 2 & 17 & FH5:35 & 1 & 9659.50 & 22,860 & 0.1341385 & $6.1328\times10^{9}$ \\
AE--pricing & $\theta=0.126303$ & 2 & 17 & FH5:27 & 1 & 19826.50 & 20,932 & 0.3039345 & $1.2724\times10^{10}$ \\
AE--power & $\theta=0.321751$ & 3 & 17 & FH5:34 & 4 & 9690.62 & 27,812 & 0.2344045 & $1.3039\times10^{10}$ \\
Grover--SAT & $\theta=0$ & 2 & 13 & RP3:8 & 0 & 210.00 & 10,934 & 0.002002 & $4.3780\times10^{7}$ \\
Grover--AES & $\theta=0$ & 2 & 13 & RP3:8 & 0 & 210.00 & 15,989 & 0.002067 & $6.6099\times10^{7}$ \\
Grover--hash & $\theta=0$ & 2 & 13 & RP3:8 & 0 & 210.00 & 15,989 & 0.001989 & $6.3604\times10^{7}$ \\
QAOA--AA8 & $\theta=0.392699$ & 3 & 15 & RP3:30 & 2 & 2751.47 & 14,764 & 0.029235 & $8.6325\times10^{8}$ \\
QAOA--KSAT12 & $\theta=0.322448$ & 3 & 15 & FH5:11 & 2 & 2424.00 & 18,366 & 0.06579 & $2.4166\times10^{9}$ \\
QAOA--MaxCut & $\theta=0.940567$ & 2 & 15 & FH5:13 & 3 & 1362.00 & 20,644 & 0.0123 & $5.0784\times10^{8}$ \\
QLSA--HHL & $\theta=1.1781$ & 2 & 13 & RP3:7 & 1 & 436.75 & 11,168 & 0.0054275 & $1.2123\times10^{8}$ \\
QLSA--QSVT & $\theta=1.0472$ & 2 & 13 & RP3:7 & 2 & 390.00 & 8,135 & 0.004979 & $8.1008\times10^{7}$ \\
QPE--factor & $\theta=0.173036$ & 2 & 19 & FH5:15 & 1 & 222814.50 & 36,781 & 3.2406495 & $2.3839\times10^{11}$ \\
QPE--state prep. & $\theta=0.141897$ & 2 & 17 & FH5:41 & 1 & 42090.00 & 36,423 & 0.50456 & $3.6755\times10^{10}$ \\
QPE--Trotter & $N=1$ & 3 & 17 & FH5:9 & 1 & 25053.50 & 26,980 & 0.370243 & $1.9978\times10^{10}$ \\
QSim--JW & $\theta=0.8$ & 2 & 13 & RP3:6 & 3 & 224.00 & 13,424 & 0.003796 & $1.0192\times10^{8}$ \\
QSim--tiled & $\theta=0.8$ & 2 & 13 & RP3:6 & 3 & 224.00 & 13,424 & 0.004004 & $1.0750\times10^{8}$ \\
QSim--Heisenberg & $\theta=0.8$ & 2 & 13 & RP3:6 & 3 & 336.00 & 15,109 & 0.002925 & $8.8388\times10^{7}$ \\
Shor--binary ECDLP & $\theta=0$ & 2 & 15 & RP3:6 & 0 & 2162.00 & 16,333 & 0.0242775 & $7.9305\times10^{8}$ \\
Shor--prime ECDLP & $\theta=0$ & 2 & 15 & RP3:6 & 0 & 2932.00 & 22,170 & 0.0326325 & $1.4469\times10^{9}$ \\
Shor--RSA15 & $\theta=0.392699$ & 2 & 15 & RP3:6 & 2 & 2252.75 & 14,088 & 0.0268275 & $7.5589\times10^{8}$ \\
\bottomrule
\end{tabular}%
}
\end{table*}

\begin{table*}[htbp]
\centering
\scriptsize
\caption{Absolute All-$T$ controls from the same corrected records. Paired entries are fast / optimized pool; the latter can retain fast. Qubits are physical, times seconds, and volumes physical-qubit--stabilizer-rounds. Candidate parameters are independently selected.}
\label{tab:matched-layout-resources-20}
\resizebox{\textwidth}{!}{%
\begin{tabular}{lrrrrr}
\toprule
Circuit & $d_f/d_o$ & Factory:count & $Q_f/Q_o$ & $t_f/t_o$ & $V_f/V_o$ \\
\midrule
AE--BHMT & 17/17 & FH5:25/FH5:25 & 20,450/20,450 & 0.424983/0.424983 & $1.7382\times10^{10}$/$1.7382\times10^{10}$ \\
AE--pricing & 17/17 & FH5:22/FH5:22 & 19,727/19,727 & 0.6965665/0.6965665 & $2.7482\times10^{10}$/$2.7482\times10^{10}$ \\
AE--power & 19/19 & FH5:26/FH5:26 & 31,501/31,501 & 1.7813925/1.7813925 & $1.1223\times10^{11}$/$1.1223\times10^{11}$ \\
Grover--SAT & 13/13 & RP3:7/RP3:8 & 12,516/10,934 & 0.002704/0.002002 & $6.7687\times10^{7}$/$4.3780\times10^{7}$ \\
Grover--AES & 13/13 & RP3:7/RP3:8 & 15,886/15,989 & 0.0026585/0.002067 & $8.4466\times10^{7}$/$6.6099\times10^{7}$ \\
Grover--hash & 13/13 & RP3:7/RP3:8 & 15,886/15,989 & 0.002574/0.001989 & $8.1781\times10^{7}$/$6.3604\times10^{7}$ \\
QAOA--AA8 & 17/17 & FH5:28/FH5:28 & 21,173/21,173 & 0.258995/0.258995 & $1.0967\times10^{10}$/$1.0967\times10^{10}$ \\
QAOA--KSAT12 & 17/17 & FH5:38/FH5:38 & 29,353/29,353 & 0.466242/0.466242 & $2.7371\times10^{10}$/$2.7371\times10^{10}$ \\
QAOA--MaxCut & 15/15 & RP3:24/RP3:25 & 18,187/17,392 & 0.0271725/0.021855 & $9.8837\times10^{8}$/$7.6020\times10^{8}$ \\
QLSA--HHL & 15/15 & RP3:11/RP3:13 & 15,052/14,809 & 0.04485/0.039285 & $1.3502\times10^{9}$/$1.1635\times10^{9}$ \\
QLSA--QSVT & 15/15 & RP3:10/RP3:12 & 12,255/10,665 & 0.0228525/0.0185175 & $5.6011\times10^{8}$/$3.9498\times10^{8}$ \\
QPE--factor & 19/19 & FH5:17/FH5:17 & 36,542/36,542 & 3.558738/3.558738 & $2.6009\times10^{11}$/$2.6009\times10^{11}$ \\
QPE--state prep. & 19/19 & FH5:28/FH5:30 & 39,193/39,675 & 0.7763305/0.7663555 & $6.0853\times10^{10}$/$6.0810\times10^{10}$ \\
QPE--Trotter & 17/17 & FH5:29/FH5:29 & 32,954/32,954 & 0.661861/0.661861 & $4.3622\times10^{10}$/$4.3622\times10^{10}$ \\
QSim--JW & 15/15 & RP3:16/RP3:16 & 17,363/17,363 & 0.019575/0.019575 & $6.7976\times10^{8}$/$6.7976\times10^{8}$ \\
QSim--tiled & 15/15 & RP3:16/RP3:16 & 17,363/17,363 & 0.0195675/0.0195675 & $6.7950\times10^{8}$/$6.7950\times10^{8}$ \\
QSim--Heisenberg & 15/15 & RP3:18/RP3:21 & 17,569/16,980 & 0.0162/0.013245 & $5.6924\times10^{8}$/$4.4980\times10^{8}$ \\
Shor--binary ECDLP & 15/15 & RP3:6/RP3:6 & 16,333/16,333 & 0.0270825/0.0242775 & $8.8468\times10^{8}$/$7.9305\times10^{8}$ \\
Shor--prime ECDLP & 15/15 & RP3:6/RP3:6 & 20,823/22,170 & 0.036225/0.0326325 & $1.5086\times10^{9}$/$1.4469\times10^{9}$ \\
Shor--RSA15 & 15/15 & RP3:11/RP3:11 & 15,052/14,603 & 0.033225/0.029835 & $1.0002\times10^{9}$/$8.7136\times10^{8}$ \\
\bottomrule
\end{tabular}%
}
\end{table*}
\begin{table*}[htbp]
\centering
\scriptsize
\caption{Space and time factors: the first three are $H_f/H_o$ (layout with Hybrid), the last three $T_f/H_o$ (combined). $G_QG_t=G_V$ before rounding. A factor below one is retained and indicates a larger resource in that dimension. All entries use the same corrected conditional model.}
\label{tab:matched-layout-features-20}
\resizebox{\textwidth}{!}{%
\begin{tabular}{lrrrrrr}
\toprule
Circuit & $G_Q$ (layout) & $G_t$ (layout) & $G_V$ (layout) & $G_Q$ (combined) & $G_t$ (combined) & $G_V$ (combined) \\
\midrule
AE--BHMT & 0.9409 & 1.0975 & 1.0327 & 0.8946 & 3.1682 & 2.8342 \\
AE--pricing & 1.0045 & 1.0399 & 1.0446 & 0.9424 & 2.2918 & 2.1599 \\
AE--power & 0.9982 & 1.1912 & 1.1890 & 1.1326 & 7.5997 & 8.6077 \\
Grover--SAT & 1.1447 & 1.3506 & 1.5461 & 1.1447 & 1.3506 & 1.5461 \\
Grover--AES & 0.9936 & 1.2862 & 1.2779 & 0.9936 & 1.2862 & 1.2779 \\
Grover--hash & 0.9936 & 1.2941 & 1.2858 & 0.9936 & 1.2941 & 1.2858 \\
QAOA--AA8 & 0.9093 & 1.2399 & 1.1274 & 1.4341 & 8.8591 & 12.7048 \\
QAOA--KSAT12 & 1.0244 & 1.0682 & 1.0943 & 1.5982 & 7.0868 & 11.3263 \\
QAOA--MaxCut & 0.9783 & 1.0994 & 1.0755 & 0.8810 & 2.2091 & 1.9462 \\
QLSA--HHL & 1.0302 & 1.1689 & 1.2041 & 1.3478 & 8.2635 & 11.1373 \\
QLSA--QSVT & 1.1657 & 1.0548 & 1.2296 & 1.5065 & 4.5898 & 6.9143 \\
QPE--factor & 1.0000 & 1.0000 & 1.0000 & 0.9935 & 1.0982 & 1.0910 \\
QPE--state prep. & 1.0000 & 1.0000 & 1.0000 & 1.0761 & 1.5386 & 1.6556 \\
QPE--Trotter & 1.0642 & 1.0162 & 1.0814 & 1.2214 & 1.7876 & 2.1835 \\
QSim--JW & 1.0000 & 1.1524 & 1.1524 & 1.2934 & 5.1567 & 6.6699 \\
QSim--tiled & 1.0000 & 1.0925 & 1.0925 & 1.2934 & 4.8870 & 6.3210 \\
QSim--Heisenberg & 0.8817 & 1.4467 & 1.2755 & 1.1628 & 5.5385 & 6.4402 \\
Shor--binary ECDLP & 1.0000 & 1.1155 & 1.1155 & 1.0000 & 1.1155 & 1.1155 \\
Shor--prime ECDLP & 0.9392 & 1.1101 & 1.0426 & 0.9392 & 1.1101 & 1.0426 \\
Shor--RSA15 & 1.0637 & 1.0663 & 1.1342 & 1.0684 & 1.2385 & 1.3232 \\
\bottomrule
\end{tabular}%
}
\end{table*}
\begin{table*}[htbp]
\centering
\scriptsize
\caption{Twelve independent QAOA checkpoints, using the same corrected catalogue and analog-error history as the main benchmarks. Paired entries are All-$T$ fast / Hybrid optimized. All-$T$ optimized retains fast, so $T_o/H_o=T_f/H_o$. These readout-aware checkpoints are not shot-weighted full-training costs. Times are seconds and volumes physical-qubit--stabilizer-rounds.}
\label{tab:qaoa-trajectory}
\resizebox{\textwidth}{!}{%
\begin{tabular}{llrrrrrrr}
\toprule
Family & Checkpoint & $d_{T_f}/d_{H_o}$ & $Q_{T_f}/Q_{H_o}$ & $t_{T_f}/t_{H_o}$ & $V_{T_f}/V_{H_o}$ & $T_f/H_f$ & $H_f/H_o$ & $T_f/H_o$ \\
\midrule
AA8 & 000 & 17/15 & 21,414/14,258 & 0.2607545/0.0305775 & $1.1168\times10^{10}$/$8.7195\times10^{8}$ & 10.0359 & 1.2762 & 12.8076 \\
AA8 & 020 & 17/15 & 21,173/14,315 & 0.2595305/0.0300525 & $1.0990\times10^{10}$/$8.6040\times10^{8}$ & 10.7368 & 1.1897 & 12.7732 \\
AA8 & 040 & 17/15 & 20,932/14,661 & 0.2598705/0.0279825 & $1.0879\times10^{10}$/$8.2050\times10^{8}$ & 10.8192 & 1.2255 & 13.2592 \\
AA8 & 060 & 17/15 & 20,932/14,661 & 0.259573/0.02865 & $1.0867\times10^{10}$/$8.4008\times10^{8}$ & 10.7619 & 1.2020 & 12.9355 \\
AA8 & 080 & 17/15 & 21,173/14,521 & 0.2558585/0.0287775 & $1.0835\times10^{10}$/$8.3576\times10^{8}$ & 11.0067 & 1.1778 & 12.9638 \\
AA8 & 100 & 17/15 & 21,173/14,764 & 0.258995/0.029235 & $1.0967\times10^{10}$/$8.6325\times10^{8}$ & 11.2689 & 1.1274 & 12.7048 \\
KSAT12 & 000 & 17/15 & 29,112/17,128 & 0.4998425/0.0401175 & $2.9103\times10^{10}$/$1.3743\times10^{9}$ & 18.7037 & 1.1322 & 21.1770 \\
KSAT12 & 020 & 17/15 & 28,630/15,884 & 0.4699395/0.0437175 & $2.6909\times10^{10}$/$1.3888\times10^{9}$ & 16.7364 & 1.1577 & 19.3753 \\
KSAT12 & 040 & 17/15 & 29,112/17,917 & 0.4912405/0.0546825 & $2.8602\times10^{10}$/$1.9595\times10^{9}$ & 12.8025 & 1.1401 & 14.5966 \\
KSAT12 & 060 & 17/15 & 30,317/18,768 & 0.4976495/0.0521325 & $3.0174\times10^{10}$/$1.9568\times10^{9}$ & 14.2121 & 1.0850 & 15.4200 \\
KSAT12 & 080 & 17/15 & 29,112/18,121 & 0.478618/0.0574725 & $2.7867\times10^{10}$/$2.0829\times10^{9}$ & 13.3789 & 1.0000 & 13.3789 \\
KSAT12 & 100 & 17/15 & 30,076/18,356 & 0.481151/0.0527625 & $2.8942\times10^{10}$/$1.9370\times10^{9}$ & 13.6228 & 1.0968 & 14.9416 \\
\bottomrule
\end{tabular}%
}
\end{table*}

\begin{figure*}[htbp]
\centering
\includegraphics[width=\textwidth]{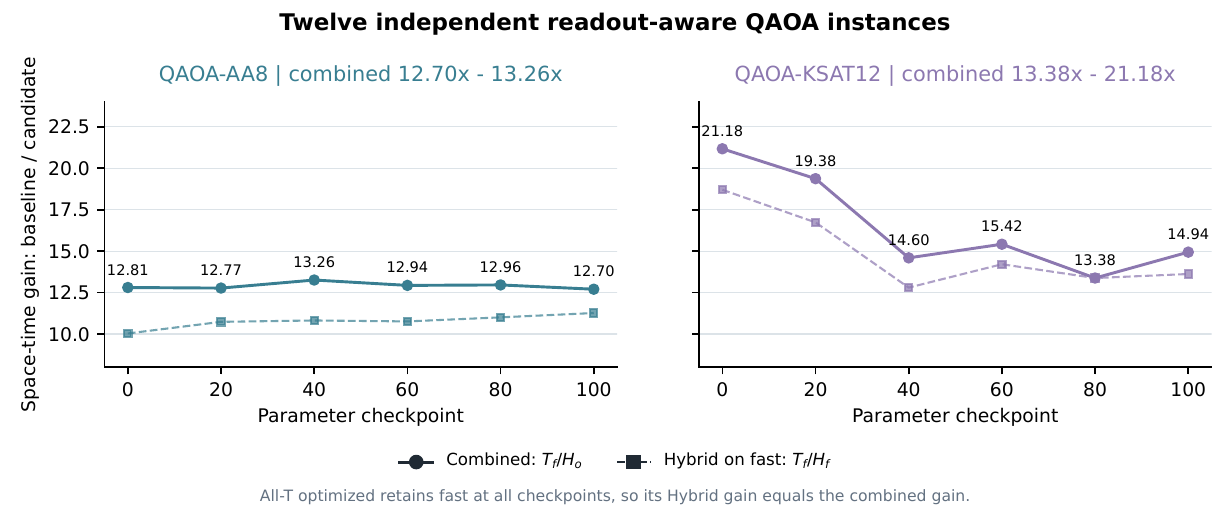}
\caption{\textbf{Hybrid and layout robustness across twelve QAOA parameter checkpoints.} Six supplied readout-aware instances are evaluated per family in all four arms. Solid curves show the combined $V(T_f)/V(H_o)$ gain; dashed curves show Hybrid alone on fast, $V(T_f)/V(H_f)$. All-$T$ optimized retains fast at every checkpoint, so $V(T_o)/V(H_o)$ equals the combined gain here. Combined gains span 12.70--13.26 for AA8 and 13.38--21.18 for KSAT12. Each arm independently selects resource parameters under the same corrected conditional mean-timing LS model. Lines guide the eye and do not estimate unsampled points. These instances are separate from the twenty main benchmarks and do not establish shot-weighted full-training cost.}
\label{fig:sm_qaoa_trajectory}
\end{figure*}

\section{Numerical reconstruction of the secp256k1 example}
\label{sec:sm_ecdlp}

\subsection{Reconstruction boundary}

The secp256k1 application-scale calculation is reconstructed hierarchically rather than expanded into a complete elementary circuit.

The strongest source-reconstructed boundary is one controlled affine point addition. Below this level, arithmetic routines, measurements, resets, and available dynamic dependencies are reconstructed from the released implementation. Above this boundary, signed-window multiplicities, table-lookups, and the semiclassical-QFT structure are taken from the paper-level construction.

\subsection{Controlled point-addition resources}

One controlled affine point addition has peak logical width
\begin{equation}
1+256+256+256+66=835
\end{equation}
logical qubits.

Its reconstructed arithmetic representation contains
\begin{equation}
N_{\mathrm{CCX}}^{\mathrm{CPA}}
=
59{,}119{,}932,
\end{equation}
together with 99,644,941 CX gates, 24,105,455 measurements, and 24,105,455 resets.

The selected local LS profiles compose to a base duration of 743,749,066 logical clocks per point addition and consume 413,839,524 $T/T^\dagger$ states. The largest individual request contains four T states. This duration is evaluated before global supply waiting; the full execution time includes the shared-inventory waiting described in the execution-time subsection below.

The 220 multi-qubit local circuits and 23 single-qubit or dynamic-event circuits supply the 1,931 arithmetic families. Each local profile retains its duration, timestamped T requests, charged patch footprint, and port contract. Ready operations are selected deterministically in their original invocation order while preserving shared-qubit and classical dependencies. Concurrent blocks satisfy
\begin{equation}
N_{\rm idle\ data}+\sum_{b\in\mathrm{active}}N_{{\rm patches},b}\leq1{,}256.
\end{equation}
Operations are delayed when their concurrent footprint would exceed this area. At block boundaries, the standard frame convention retains charged terminal H operations; cross-block H cancellation is not applied. Pauli-frame operations and measurement/reset/feed-forward events carry zero additional time in this model, while a surviving physical H costs three clocks.

Upper-level calls retain the normalized ordered composition, including the resource-equivalent inverse-EEA and abstract MCX profiles. Parent contributions available only as counts use serial primitive profiles. Table lookups retain their serial CCX composition, evaluated with the same local CCX profile. These ordering and timing conventions define the hierarchical estimate.

\subsection{Top-level algorithmic composition}

For $n=256$ and signed-window size $w=16$, the model contains 28 signed-window stages. Each stage contains one controlled affine point addition and five sequential table lookups. Each table lookup contributes $2^{16}$ Toffoli gates.

The full Toffoli count is therefore
\begin{equation}
\begin{aligned}
N_{\mathrm{CCX}}^{\mathrm{full}}
&=
28
\left(
59{,}119{,}932
+
5\times2^{16}
\right)
\\
&=
1{,}664{,}533{,}136.
\end{aligned}
\label{eq:sm_ecdlp_toffoli}
\end{equation}

Using seven $T$ states per Toffoli gives
\begin{equation}
7N_{\mathrm{CCX}}^{\mathrm{full}}
=
11{,}651{,}731{,}952.
\end{equation}

The semiclassical-QFT part retains a fixed synthesis model with 513 nontrivial rotations, each assigned a 67-$T$ cost,
\begin{equation}
513\times67=34{,}371.
\end{equation}

The final modeled demand is therefore
\begin{equation}
N_T^{\mathrm{full}}
=
11{,}651{,}766{,}323.
\end{equation}

The fixed QFT profile contains $2{,}055+9\times34{,}371=311{,}394$ base clocks and an additive synthesis-error allocation of $10^{-4}$. Its supply waits are evaluated using the inventory arriving from the preceding windows.

\subsection{Code distance and physical footprint}

The peak width of 835 logical qubits is represented using the compact-block convention
\begin{equation}
N_{\mathrm{patch}}
=
\left\lceil
1.5N_{\mathrm{logical}}
\right\rceil
+3,
\end{equation}
giving
\begin{equation}
N_{\mathrm{patch}}
=
1{,}256.
\end{equation}

For each fixed local-layout selection, we jointly search the odd distances $d=5,7,\ldots,37$, catalogue factories supporting $p_{\mathrm{phys}}=10^{-3}$, buffer capacities
\begin{equation}
B\in\{B_0,2B_0,4B_0,8B_0\},\qquad B_0=\max(835,n_{T,\max}),
\end{equation}
and factory multiplicities $N_F=1,\ldots,B$. Here $n_{T,\max}=4$ is the largest individual request, so $B_0=835$. Each local layout remains fixed during this resource search. We select the feasible candidate with the lowest provisioned space--time volume, breaking ties by runtime, footprint, and a fixed configuration order.

One shared buffer starts full at the beginning of the entire computation. Its initial preparation time is excluded under the pre-supply convention; all buffer patches are charged for the complete execution, including empty slots. Factories produce synchronized deterministic batches of $N_F$ states every $c_F$ clocks, with excess production discarded at capacity. A request with insufficient inventory waits for the necessary batches, shifting subsequent execution. Both inventory and production phase persist across blocks, families, point additions, and table lookups.

The selected configuration has $B=835$ and $N_F=58$. The buffer is at the lower boundary of the declared search set. 

At $p_{\mathrm{phys}}=10^{-3}$, the selected code distance is
\begin{equation}
d=35.
\end{equation}

Each grid-resident patch therefore contains
\begin{equation}
2d^2-1
=
2\times35^2-1
=
2{,}449
\end{equation}
physical qubits.

The data block requires
\begin{equation}
N_{\mathrm{data}}
=
1{,}256\times2{,}449
=
3{,}075{,}944
\end{equation}
physical qubits.

The buffer contributes
\begin{equation}
N_{\rm buffer}=835\times2{,}449=2{,}044{,}915
\end{equation}
physical qubits. The selected T-state plan contains 58 factories, each with physical footprint 39,100 qubits,
\begin{equation}
N_{\mathrm{factory}}
=
58\times39{,}100
=
2{,}267{,}800.
\end{equation}

The peak provisioned physical footprint is
\begin{equation}
\begin{aligned}
N_{\mathrm{phys}}
&=(1{,}256+B)(2d^2-1)+N_FQ_F\\
&=3{,}075{,}944+2{,}044{,}915+2{,}267{,}800\\
&=7{,}388{,}659.
\end{aligned}
\label{eq:sm_ecdlp_physical}
\end{equation}

The chosen high-fidelity distillation factory has $p_T^{\rm out}=3.3\times10^{-14}$ and $c_F=97.5$ clocks. Its catalogue footprint $Q_F=39{,}100$ is already a physical-qubit count. The separate catalogue quantity $L_F=264$ enters the protected-patch exposure proxy. With the fixed QFT synthesis allocation, the error contributions are
\begin{align}
P_{\rm NC}&=N_T^{\rm full}p_T^{\rm out}+10^{-4}\notag\\
&\simeq4.845082887\times10^{-4},\\
P_{\rm SC}&=(1{,}256+B+N_FL_F)C_{\rm exec}d\,p_L\notag\\
&\simeq1.302236208\times10^{-3}.
\end{align}
Candidates are screened at $P_{\rm SC}\leq0.009$ and verified against the final surface-code budget of 0.01; the non-Clifford budget is also 0.01. For this complete T demand, the accepted-state error test excludes the catalogue cultivation entries. The selected plan therefore uses distillation throughout.

\subsection{Execution time}

Using\begin{equation}\tau_{\mathrm{round}}=500~\mathrm{ns}\end{equation}and $d=35$, the logical-clock duration is\begin{equation}\tau_{\mathrm{clock}}=35\times500~\mathrm{ns}=17.5~\mu\mathrm{s}.\end{equation}
The execution occupies $C_{\rm exec}d=748{,}282{,}599{,}400$ stabilizer rounds, giving\begin{align}T_{\mathrm{exec}}&=374{,}141.299700~\mathrm{s}\\&\simeq103.9281~\mathrm{h}\simeq4.3303~\mathrm{days}.\end{align}Charging the provisioned footprint throughout this execution gives\begin{equation}V_{\rm ST}=N_{\rm phys}C_{\rm exec}d\simeq5.528804963\times10^{18}\end{equation}physical-qubit--stabilizer-rounds.
The result corresponds to one execution with the stated local profiles, global supply model, and upper-level composition. The data block, buffer, and factory fleet are reused across sequential stages and are counted by peak provisioned footprint. 

\section{Additional technical parameters and reference tables}
\label{sec:sm_tables}

This section collects auxiliary numerical assumptions for the explicit-benchmark resource model. The secp256k1 reconstruction uses selected local LS profiles with the hierarchical scheduling and continuous-inventory assumptions specified in Supplementary Sec.~\ref{sec:sm_ecdlp}.

\subsection{Decoder and feedback timing}

The end-to-end feedback latency is decomposed as
\begin{equation}
\tau_{\mathrm{rt}}
=
\tau_{\mathrm{in}}
+
\tau_{\mathrm{decode}}
+
\tau_{\mathrm{out}}.
\end{equation}

Published measurements report different subsets of these terms, so decoder-core runtime and end-to-end control latency should not be compared directly. The numerical baseline in this work therefore uses an effective stabilizer-round duration of 500 ns rather than assigning a universal decoder latency.


\subsection{Distillation and cultivation catalogue}

The discrete resource-state catalogue used by the optimizer specifies, for each protocol, the applicable physical-error regime, accepted-state output error, physical footprint, discard probability when relevant, and output cadence.

Supplementary Tables~\ref{tab:factory-distillation-catalogue} and
\ref{tab:factory-cultivation-catalogue} list the discrete factory
catalogue used by the estimator. Entries are filtered by
$p_{\mathrm{phys}}$ and output fidelity before multiplicity is sized.

\begin{table*}[htbp]
\centering
\scriptsize

\caption{Distillation protocols in the estimator catalogue. $Q_F$ is
the absolute physical-qubit footprint of one factory. The listed cadence
parameters are interpreted as logical clocks per output by this estimator;
this is the retained catalogue timing convention, not a uniform calibration
of the source protocols.}
\label{tab:factory-distillation-catalogue}

\resizebox{0.5\textwidth}{!}{%
\begin{tabular}{lrrrr}
\toprule
Protocol & $p_{\mathrm{phys}}$ & $p_T^{\mathrm{out}}$ & $Q_F$ & Cadence \\
\midrule
$(15\text{-to-}1)*{7,3,3}$ & $10^{-4}$ & $4.4\times10^{-8}$ & 810 & 18.1 \\
$(15\text{-to-}1)*{9,3,3}$ & $10^{-4}$ & $9.3\times10^{-10}$ & 1,150 & 18.1 \\
$(15\text{-to-}1)*{11,5,5}$ & $10^{-4}$ & $1.9\times10^{-11}$ & 2,070 & 30.0 \\
$(15\text{-to-}1)^4\times(20\text{-to-}4)$ & $10^{-4}$ & $2.4\times10^{-15}$ & 16,400 & 90.3 \\
$(15\text{-to-}1)^4\times(15\text{-to-}1)$ & $10^{-4}$ & $6.3\times10^{-25}$ & 18,600 & 67.8 \\
$(15\text{-to-}1)*{17,7,7}$ & $10^{-3}$ & $4.5\times10^{-8}$ & 4,620 & 42.6 \\
$(15\text{-to-}1)^6\times(20\text{-to-}4)$ & $10^{-3}$ & $1.4\times10^{-10}$ & 43,300 & 130 \\
$(15\text{-to-}1)^4\times(20\text{-to-}4)$ & $10^{-3}$ & $2.6\times10^{-11}$ & 46,800 & 157 \\
$(15\text{-to-}1)^6\times(15\text{-to-}1)$ (low) & $10^{-3}$ & $2.7\times10^{-12}$ & 30,700 & 82.5 \\
$(15\text{-to-}1)^6\times(15\text{-to-}1)$ (high) & $10^{-3}$ & $3.3\times10^{-14}$ & 39,100 & 97.5 \\
\bottomrule
\end{tabular}%
}

\vspace{1.2\baselineskip}

\caption{Cultivation catalogue inputs. Discard is the overall single-candidate rate; the listed cadence is the expected single-candidate period in logical clocks at data distance $d$, before the PPC rescaling in Supplementary Sec.~\ref{subsec:sm_ppc}. $Q_F$ is retained as the catalogue's absolute factory footprint, independently of $D_1,D_2$ and data $d$. The RP$^2$ $f=3$ footprint is the corrected maximum-active-qubit input. Cadences and output errors remain declared catalogue scenario values: protocol time units and proxy-state error conventions have not been uniformly hardware-calibrated. The $p_{\rm phys}=10^{-4}$ Gidney entry is an ungrown fallback without a matched PPC correction.}
\label{tab:factory-cultivation-catalogue}

\resizebox{0.8\textwidth}{!}{%
\begin{tabular}{llrrrrr}
\toprule
Source & Protocol & $p_{\mathrm{phys}}$ & $p_T^{\mathrm{out}}$ &
Discard & $Q_F$ & Cadence \\
\midrule
Gidney MSC & $D_1=3$, graft $D_2=15$ & $10^{-3}$ & $3.0\times10^{-6}$ & 0.80 & 337 & $555/d$ \\
Gidney MSC & $D_1=5$, graft $D_2=15$ & $10^{-3}$ & $2.0\times10^{-9}$ & 0.99 & 449 & $21900/d$ \\
Gidney MSC & fault distance 5, ungrown & $10^{-4}$ & $6.0\times10^{-15}$ & 0.20 & 449 & $273.75/d$ \\
Folded-H MSC & $f=3$, grow $d=3\to13$ & $10^{-3}$ & $1.76\times10^{-6}$ & 0.535 & 169 & $146.2/d$ \\
Folded-H MSC & $f=5$, grow $d=5\to15$ & $10^{-3}$ & $9.28\times10^{-10}$ & 0.878 & 241 & $950.82/d$ \\
RP$^2$ MSC & $f=3$, graft $d=7$ & $10^{-3}$ & $1.5\times10^{-6}$ & 0.58 & 103 & $178.57/d$ \\
RP$^2$ MSC & $f=5$, graft $d=11$ & $10^{-3}$ & $1.0\times10^{-9}$ & 0.934 & 251 & $2242.42/d$ \\
Claes MSC & $f=3$, graft $d=7$ & $10^{-3}$ & $1.0\times10^{-6}$ & 0.66 & 337 & $252.94/d$ \\
Claes MSC & $f=5$, graft $d=11$ & $10^{-3}$ & $2.0\times10^{-9}$ & 0.99 & 449 & $16400/d$ \\
\bottomrule
\end{tabular}%
}

\end{table*}

\clearpage

\subsection{Physical-resource bookkeeping}

Grid-resident data, routing/ancilla, and direct-rotation patches use
\begin{equation}
Q_{\mathrm{patch}}
=
2d^2-1.
\end{equation}

Factory catalogue entries are already absolute physical-qubit counts. Thus
\begin{equation}
N_{\mathrm{phys}}
=
N_{\mathrm{grid}}
(2d^2-1)
+
N_FQ_F.
\end{equation}

Factory-access interfaces in a layout are not complete factory bodies. The selected catalogue footprint is charged for each factory; additional buffers, external delivery routing, and a closed processor floorplan are not independently constructed or validated by this resource model. The surface-error proxy uses the catalogue logical-patch count $L_F$ separately from this absolute footprint.

\nocite{Gidney2023inplaceY,griffiths1996semiclassical,
luo2026quantumalgorithmellipticcurve}
\bibliographystyle{unsrt}
\bibliography{ref}

\end{document}